\documentclass[english]{article}
\usepackage[T1]{fontenc}
\usepackage[latin9]{inputenc}
\usepackage{geometry}
\usepackage{mathtools}
\usepackage{amsmath}
\usepackage{amsthm}
\usepackage{amssymb}
\usepackage{graphicx}

\makeatletter
\numberwithin{equation}{section}
\newcommand{\address}[1]{
	\par {\raggedright #1
	\vspace{1.4em}
	\noindent\par}
}
\theoremstyle{plain}
\newtheorem{thm}{\protect\theoremname}
\theoremstyle{plain}
\newtheorem{prop}[thm]{\protect\propositionname}
\theoremstyle{remark}
\newtheorem*{rem*}{\protect\remarkname}
\theoremstyle{plain}
\newtheorem{lem}[thm]{\protect\lemmaname}
\theoremstyle{plain}
\newtheorem{cor}[thm]{\protect\corollaryname}

\usepackage{algorithm,algpseudocode}

\ifdefined\showcaptionsetup
 \PassOptionsToPackage{caption=false}{subfig}
\fi
\usepackage{subfig}
\makeatother

\usepackage{babel}
\providecommand{\corollaryname}{Corollary}
\providecommand{\lemmaname}{Lemma}
\providecommand{\propositionname}{Proposition}
\providecommand{\remarkname}{Remark}
\providecommand{\theoremname}{Theorem}

\begin{document}
\title{Oscillatory solutions at the continuum limit of Lorenz 96 systems}
\author{Di Qi\textsuperscript{a} and Jian-Guo Liu\textsuperscript{b} }
\maketitle

\address{\textsuperscript{a}Department of Mathematics, Purdue University,
150 North University Street, West Lafayette, IN 47907, USA}

\address{\textsuperscript{b}Department of Mathematics and Department of Physics,
Duke University, Durham, NC 27708, USA}
\begin{abstract}
In this paper, we study the generation and propagation of oscillatory
solutions observed in the widely used Lorenz 96 (L96) systems. First, period-two
oscillations between adjacent grid points are found in the leading-order
expansions of the discrete L96 system. The evolution of the envelope
of period-two oscillations is described by a set of modulation equations
with strictly hyperbolic structure. The modulation equations are found
to be also subject to an additional reaction term dependent on the
grid size, and the period-two oscillations will break down into fully
chaotic dynamics when the oscillation amplitude grows large. Then,
similar oscillation solutions are analyzed in the two-layer L96 model
including multiscale coupling. Modulation equations for period-three
oscillations are derived based on a weakly nonlinear analysis in the
transition between oscillatory and nonoscillatory regions. Detailed
numerical experiments are shown to confirm the analytical results.
\end{abstract}

\section{Introduction}

The Lorenz 96 (L96) system is a simple and illustrative model designed
by E. N. Lorenz in 1996 \cite{lorenz1996proceedings} to study
various representative features observed in the atmosphere. The original
L96 model is later generalized to a two-layer version \cite{wilks2005effects,arnold2013stochastic}
to include multiscale interactions. Though the L96 equations are deterministic,
they demonstrate intrinsically chaotic behaviors in direct numerical
solutions that display large uncertainty and instability \cite{orrell2003model,majda2016introduction}.
It shows that the L96 models can produce many remarkable statistical
and stochastic features \cite{lorenz1998optimal,olbers2001gallery,abramov2004quantifying}
in common with the climate system while maintain a much cleaner mathematical
setting. Furthermore, the L96 system has been widely used as a prototype
model to test model reduction methods in uncertainty quantification
and data assimilation \cite{majda2018strategies,majda2019linear,qi2023high,qi2023random},
and to analyze different aspects of multiscale stochastic behaviors
in chaotic dynamics \cite{fatkullin2004computational,karimi2010extensive,bedrossian2022regularity}.
In general, such complex chaotic behaviors in discrete dynamical systems
can be understood by properties of numerical schemes and the corresponding conservation
properties \cite{stuart1998dynamical}. However, there is still not
a thorough study from this perspective in the context of L96 systems
to our knowledge.

The chaotic behavior in the L96 solutions is found to be closely linked
to the automatic generation of oscillating solutions happening at
the grid scale, which can be compared to the discrete dispersive
numerical schemes. Similar oscillating solutions on mesh scale after
the formation of shocks are generally observed and systematically
analyzed under various dispersive schemes \cite{lax1983small,lax1986dispersive,hou1991dispersive}.
In contrast to the strong convergence and stability of viscous solutions
to smooth inviscid flows such as the detailed studies in \cite{xin1992linearized,liu1993nonlinear,liu1993stability},
the oscillations generated in dispersive numerical schemes do not
vanish and are maintained in finite amplitude, while the wavelength
of the oscillations remains within the grid size. It is demonstrated
from different numerical schemes on the Burgers-Hopf equation \cite{goodman1988dispersive,levermore1996large}
that the oscillations will persist in the weak convergence of the
oscillatory approximations. Inspired by this observation in dispersive
schemes, it is found that solutions of the L96 systems exhibit similar
behaviors in its way to develop fully chaotic dynamics from smooth
initial data. 

In this paper, we study the well-known complex chaotic behaviors observed
in the discrete one and two-layer L96 systems in analog to the oscillatory
solutions in dispersive schemes. First, we treat the discrete inviscid
L96 model as a finite difference approximation and study its continuum
limit with small grid size as $h\rightarrow0$. It shows that the
L96 system agrees with the solution of the Burgers-Hopf equation in
its leading order, while the higher-order corrections make the important
contribution to create the competing oscillatory features found in
the discrete L96 solutions (Proposition \ref{prop:stable_state} and
\ref{prop:negativity}). Based on typical observations from numerical
simulations, we perform a detailed investigation about the development
of oscillations at the discrete grid size from classical smooth solutions
of the initial value problem. In particular, we find the existence
of representative period-two oscillatory solutions \cite{levermore1996large}
due to the local conservation laws maintained in the L96 schemes.
Corresponding modulation equations that describe the evolution of
an envelope of the period-two oscillations are derived to characterize
the development and evolution of these typical oscillating solutions.
The Strang-type analysis \cite{strang1964accurate} can be applied
to show the convergence of the L96 scheme (Theorem \ref{thm:estim_peri2}
and Corollary \ref{cor:converg_high}). Further, the breakdown of
the period-two oscillations due to the strong effect of an additional
reaction term indicates the generation of fully chaotic behavior in
the solution. This provides a precise characterization for the route
to chaotic solutions through the intermediate oscillating regions
in the L96 models with a finite grid size $h$.

As a further development, we seek a closer look near the transition
region from non-oscillatory to oscillatory solutions. It shows that
the solution may generate more complicated period-three structures.
Using a weakly nonlinear asymptotic analysis, we derive the modulation
equations that govern such period-three oscillation features and show
the convergence of the discrete model to this typical period-three
structure in leading order (Theorem \ref{thm:estim_peri3}). In particular,
we demonstrate the multiscale performance in the two-layer L96 model,
and show the potential period-three oscillation phenomena. It is found
that the large-scale states create a contact discontinuity in the
small-scale variable from the large-scale forcing, which induces the oscillatory
solutions. Besides, all the analytical results are supported by detailed
numerical simulations of the one-layer and two-layer L96 models with
different initial conditions.

In the rest of the paper is organized as follows: we provide a general
discussion on the L96 model and its leading-order asymptotic equations
in Section \ref{sec:Leading-order-equations}. The creation of period-two
oscillations and the corresponding modulation equations are then derived
in Section \ref{sec:Generation-of-period-two}. The large and small
scale coupling mechanism in the two-layer L96 model is discussed in
Section \ref{sec:The-two-layer-Lorenz} where a typical period-three
solution is derived. Finally, a summarizing discussion is given in
Section \ref{sec:Summary}.

\section{Leading-order equations of the Lorenz 96 model\protect\label{sec:Leading-order-equations}}

\begin{figure}
\subfloat{\centering{}\includegraphics[scale=0.45]{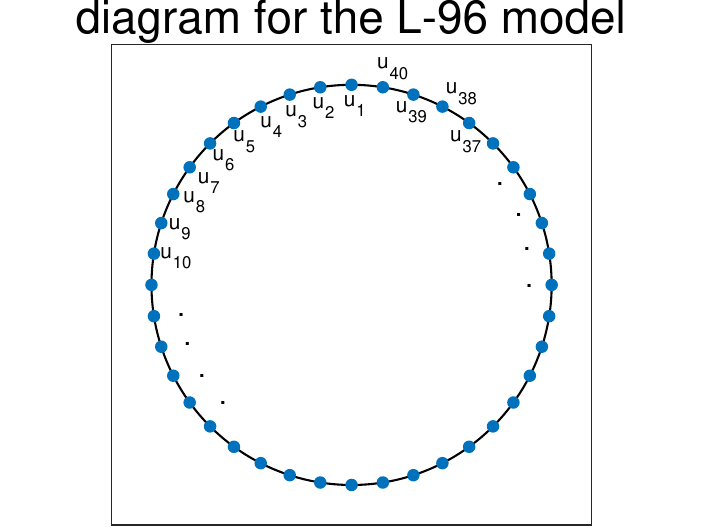}}\subfloat{\centering{}\includegraphics[scale=0.45]{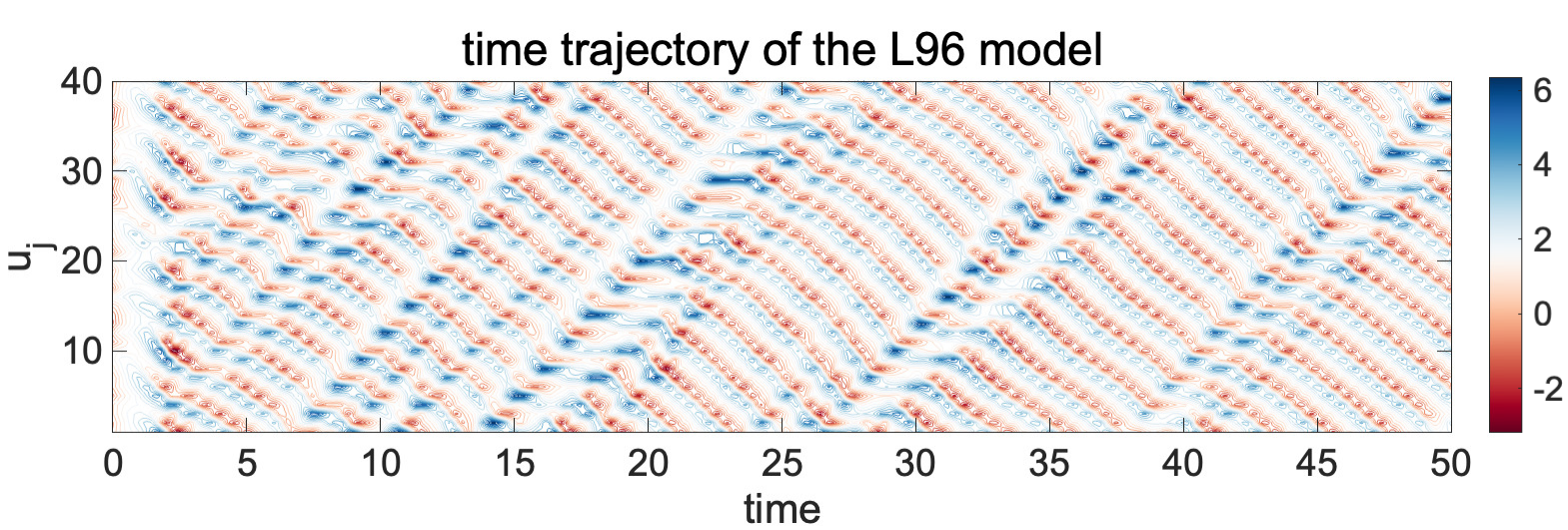}}

\caption{Illustration of the L96 model (\ref{eq:L96}) and the typical solution
with $J=40$ and $F=4$.\protect\label{fig:Illustration-l96}}
\end{figure}

The standard L96 model is given by the spatially discrete system with
uniform forcing $F$ and a linear damping
\begin{equation}
\frac{du_{j}}{dt}=\left(u_{j+1}-u_{j-2}\right)u_{j-1}-u_{j}+F,\;j=1,\cdots,J.\label{eq:L96}
\end{equation}
The model state variables $\mathbf{u}=\left(u_{1},u_{2},...,u_{J}\right)\in\mathbb{R}^{J}$
are defined with periodic boundary condition $u_{J+1}=u_{1}$ mimicking
geophysical waves in the mid-latitude atmosphere. The discrete grid
size is usually set to be $J=40$ denoting the non-dimensional midlatitude
Rossby radius \cite{abramov2004quantifying}. The model structure
and a typical solution of \eqref{eq:L96} are illustrated in Figure
\ref{fig:Illustration-l96}. The discrete solution $\mathbf{u}$ is
shown to demonstrate various representative chaotic dynamical features
\cite{lorenz1998optimal} such as westward (that is, moving to the
left) wave packages and a rapid transition from regular initial state
to highly chaotic solutions.

\subsection{The L96 system as a finite difference discretization\protect\label{subsec:The-L96-system}}

In this paper, we focus on the nonlinear coupling term in the L96
system. If the  forcing and damping terms in \eqref{eq:L96} are
set to zero, we can rewrite the\emph{ inviscid L96 model} by introducing
scaling parameters $a,h$ as
\begin{equation}
\frac{du_{j}}{dt}=\frac{1}{ah}\left(u_{j+1}-u_{j-2}\right)u_{j-1},\label{eq:L96-invisid}
\end{equation}
Above, the equation \eqref{eq:L96-invisid} is viewed as a semi-discrete
difference scheme from a continuous function $u_{j}\left(t\right)=u\left(x_{j},t\right)$
with the spatial discretization $h=x_{j+1}-x_{j}=\frac{L}{J}$ where
$J$ is the number of grid points and $L$ the domain size. For convenience
in \eqref{eq:L96-invisid}, we pick the difference factor $a=3$.
The inviscid L96 model builds an interesting bridge between the discrete
L96 equation and the continuous Burgers-Hopf equation
\begin{equation}
\partial_{t}u-u\partial_{x}u=0.\label{eq:BH}
\end{equation}
 It shows that they share very similar equilibrium statistical formalism
through a detailed statistical performance analysis in \cite{majda2000remarkable,abramov2004quantifying}.

For a closer look at the link between the equations, we introduce
the corresponding continuous state $u\left(x_{},t\right)$ defined
on $\left[0,L\right]$ with periodic boundary condition $u\left(x_{}\right)=u\left(x+L\right)$.
The continuum limit is reach as $h=\frac{L}{J}\rightarrow0$. By taking
Taylor expansion as the finite difference approximation of $u\left(x_{j},t\right)$
with $x_{j}=jh$ directly, the continuum limit of the inviscid
L96 system with $a=3$ yields
\begin{equation}
\begin{aligned}\partial_{t}u= & u\partial_{x}u-\frac{h}{2}\left(u\partial_{xx}u+2\left(\partial_{x}u\right)^{2}\right)+\frac{h^{2}}{2}\left(u\partial_{xxx}u+2\partial_{x}u\partial_{xx}u\right)\\
 & -\frac{h^{3}}{24}\left(5u\partial_{xxxx}u+16\partial_{x}u\partial_{xxx}u+6\left(\partial_{xx}u\right)^{2}\right)+O\left(h^{4}\right)\\
= & \frac{1}{2}\partial_{x}\left(u^{2}\right)-\frac{h}{2}\left(\partial_{xx}\left(u^{2}\right)-u\partial_{xx}u\right)+\frac{h^{2}}{6}\left(\partial_{xxx}\left(u^{2}\right)+u\partial_{xxx}u\right)\\
 & -\frac{h^{3}}{24}\left(\partial_{xxxx}\left(u^{2}\right)+8\partial_{x}\left(u\partial_{xxx}u\right)-5u\partial_{xxxx}u\right)+O\left(h^{4}\right).
\end{aligned}
\label{eq:L96_conti}
\end{equation}
From \eqref{eq:L96_conti}, it shows that the inviscid L96 model \eqref{eq:L96-invisid}
can be viewed as a finite difference approximation to the Burgers-Hopf
equation \eqref{eq:BH} in the leading order. Immediately, we find
the two major conservation equations in the leading order
\begin{equation}
\begin{aligned}\frac{d}{dt}\int_{0}^{L}u\:dx= & -\frac{h}{2}\int_{0}^{L}\left(\partial_{x}u\right)^{2}dx+\frac{5h^{3}}{24}\int_{0}^{L}\left(\partial_{xx}u\right)^{2}dx+O\left(h^{4}\right),\\
\frac{d}{dt}\int_{0}^{L}u^{2}dx= & \;O\left(h^{4}\right).
\end{aligned}
\label{eq:conservation}
\end{equation}
The first equation in \eqref{eq:conservation} shows that the spatially
averaged mean state $\bar{u}=\int u\:dx$ is damped by $\int\left|\partial_{x}u\right|^{2}dx$
in the next order $O\left(h\right)$. This implies the nonlinear coupling
between the mean state and subscale modes. On the other hand, the
total energy of the system, $E=\int u^{2}dx$, is conserved up to
the fourth order.

It is well-known that the initial value problem of the Burgers-Hopf
equation \eqref{eq:BH} can create an infinite derivative in finite
time due to the development of shocks. This process can be viewed
as the `cascade of energy' in turbulence from the mean state $\bar{u}=\int udx$
to multiscale fluctuation modes. To have a better understanding of
the generation of chaotic behavior as shown in Figure \ref{fig:Illustration-l96},
we need to dive into the next order approximation for the detailed
nonlinear coupling between mean and fluctuation modes.

\subsection{Conserved quantities in first-order approximation}

Next, we discuss the additional effects induced from the first-order
correction in the asymptotic approximation \eqref{eq:L96_conti}.
When the solution $u$ of the continuum equation (\ref{eq:L96_conti})
remains $C^{2}$, we can first identify the contribution from the
first-order term $O\left(h\right)$ in the asymptotic expansion and
focus on its leading order effect. This leads to the PDE
\begin{equation}
\partial_{t}u=u\partial_{xx}u-\partial_{xx}\left(u^{2}\right)=-u\partial_{xx}u-2\left(\partial_{x}u\right)^{2}.\label{eq:eqn_1st}
\end{equation}
Above in \eqref{eq:eqn_1st}, we neglect the dependence on $h$ for
the moment to focus on the individual contribution in this order.
Similar to the conservation laws for the full equation \eqref{eq:conservation},
we can find the conserved energy and a dissipation on the momentum
for the separate first-order equation (\ref{eq:eqn_1st})
\begin{align*}
\frac{d}{dt}\int udx & =-\int\left(\partial_{x}u\right)^{2}dx,\\
\frac{d}{dt}\int u^{2}dx & =0.
\end{align*}
Based on the above conservation equations, we can introduce the decomposition
for the model state $u\left(x,t\right)=\bar{u}\left(t\right)+u^{\prime}\left(x,t\right)$
into a spatial mean $\bar{u}$ and fluctuations $\int u^{\prime}dx=0$
with the following equations
\begin{equation}
\frac{d\bar{u}}{dt}=-\int\left(\partial_{x}u^{\prime}\right)^{2}dx,\quad\frac{d}{dt}\left(\bar{u}^{2}+\int u^{\prime2}dx\right)=0.\label{eq:conserv0}
\end{equation}
The conservation equations \eqref{eq:conserv0} provide a crude illustration
of the nonlinear coupling mechanism between the mean and fluctuations
in the first level: the energy in the mean will be damped by the generation
of oscillating fluctuation modes due to $u_{x}^{\prime}$ (developed
from the leading-order Burgers-Hopf); and inversely the decrease in
the mean energy will reinforce the energy in the fluctuation modes.
This corresponds to the generation of oscillation solutions approximating
the asymptotic equation \eqref{eq:L96_conti}, while the oscillating
amplitude will not vanish as $h\rightarrow0$.

Furthermore, we can derive the conservation equation for any arbitrary
function $G\left(u\right)$
\begin{equation}
\frac{d}{dt}\int G\left(u\right)dx=\int\left[G^{\prime\prime}\left(u\right)u-G^{\prime}\left(u\right)\right]\left(u_{x}\right)^{2}dx.\label{eq:conserv1}
\end{equation}
It can be checked that the above two conservation equations with $G\left(u\right)=u,u^{2}$
fit into the more general conservation equation (\ref{eq:conserv1}).
Further by setting $G\left(u\right)=\left|u\right|^{p}$, we can find
a sequence of conservation equations for any $p$
\begin{equation}
\frac{d}{dt}\int\left|u\right|^{p}dx=p\left(p-2\right)\int\mathrm{sign}\left(u\right)\left|u\right|^{p-1}\left(u_{x}\right)^{2}dx.\label{eq:conserv3}
\end{equation}
Using the above conservation equations, we are able to discover basic
properties in the solutions of the first-order equation (\ref{eq:eqn_1st}).
First, it shows that the only stable steady state solution will be
a constant with negative value. Further, we can show that the solutions
of the equation (\ref{eq:eqn_1st}) preserves negativity. More precisely,
we summarize the results in the following propositions.
\begin{prop}
\label{prop:stable_state}The only stable steady state solution of
the equation (\ref{eq:eqn_1st}) is the negative constant solution
$u\equiv a<0$.
\end{prop}

\begin{proof}
By letting $p=1$ in (\ref{eq:conserv3}), we have
\begin{equation}
\frac{d}{dt}\int\left|u\right|dx=\int_{u<0}\left(u_{x}^{\prime}\right)^{2}dx-\int_{u>0}\left(u_{x}^{\prime}\right)^{2}dx.\label{eq:ene_abs}
\end{equation}
In addition, assuming a steady state solution exist, the conditions
in (\ref{eq:conserv0}) requires
\[
\frac{d\bar{u}}{dt}=0,\;\int\left(u_{x}^{\prime}\right)^{2}dx=0\;\Rightarrow\;u_{x}^{\prime}=0\;\Rightarrow u\equiv a=\mathrm{const.}
\]
Next, consider any small mean-zero perturbations $u^{\prime}$ to
the steady state solution $u=a+u^{\prime}$. If the steady state satisfies
$a<0$ so that $u=-\left|u\right|<0$, we have $\bar{u}^{2}+\int u^{\prime2}=C$
is conserved and $\bar{u}=-\int\left|u\right|$ increases in time
according to \eqref{eq:ene_abs}. Thus the fluctuation $\int u^{\prime2}$
will decrease to return to the constant steady state. On the other
hand, if the steady state is $a>0$, the conservation relation implies
that $\bar{u}$ will decrease and the fluctuations $u^{\prime}$ will
increase due to the equations for $\bar{u}$ and $\int\left|u\right|$.
Then the solution diverges from the original steady state.
\end{proof}
\begin{prop}
\label{prop:negativity}If the initial value of the equation (\ref{eq:eqn_1st})
is fully negative, that is, $\max u_{0}\left(x\right)=b<0$, the solution
will remain negative for the entire time $t>0$
\begin{equation}
\max_{x}u\left(x,t\right)\leq b<0.\label{eq:bnd_u}
\end{equation}
\end{prop}

\begin{proof}
Let
\[
G\left(u\right)=\begin{cases}
\left(u-b\right)^{2}, & u\geq b,\\
0, & u<b.
\end{cases}
\]
We have $G^{\prime\prime}u-G^{\prime}=2b$ when $u\geq b$ and $0$
otherwise. The conservation equation (\ref{eq:conserv1}) gives that
$\int G\left(u\right)$ is decreasing since
\[
\frac{d}{dt}\int G\left(u\right)dx=2b\int_{u\geq b}\left(u_{x}\right)^{2}dx<0.
\]
Together with the initial maximum value $b$ and the definition of
the function $G$, we find for all the time $t>0$
\[
\int_{u\geq b}\left(u-b\right)^{2}dx=\int G\left(u\right)dx\leq\int G\left(u_{0}\right)dx=0\;\Rightarrow\;\int_{u\geq b}\left(u-b\right)^{2}dx\equiv0.
\]
This implies for all $t>0$ and almost everywhere in $x$
\[
u\left(x,t\right)\leq b<0.
\]
\end{proof}
On the other hand, there is no similar result for the positive solutions
of (\ref{eq:eqn_1st}). In fact from the proof in Proposition \ref{prop:negativity},
a positive mean state $\bar{u}$ will excite more small-scale fluctuation
modes. This will induce energy cascade to small scales in the transient
state and drive the solution away from the positive initial state
finally to the negative regime.

\begin{figure}
\includegraphics[scale=0.35]{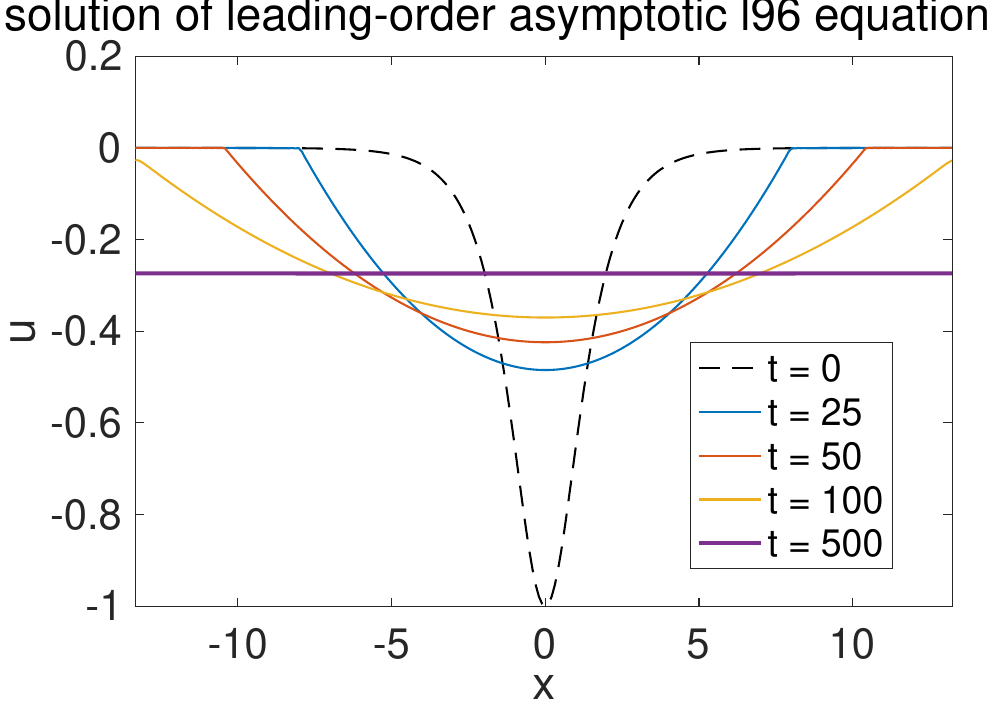}\includegraphics[scale=0.35]{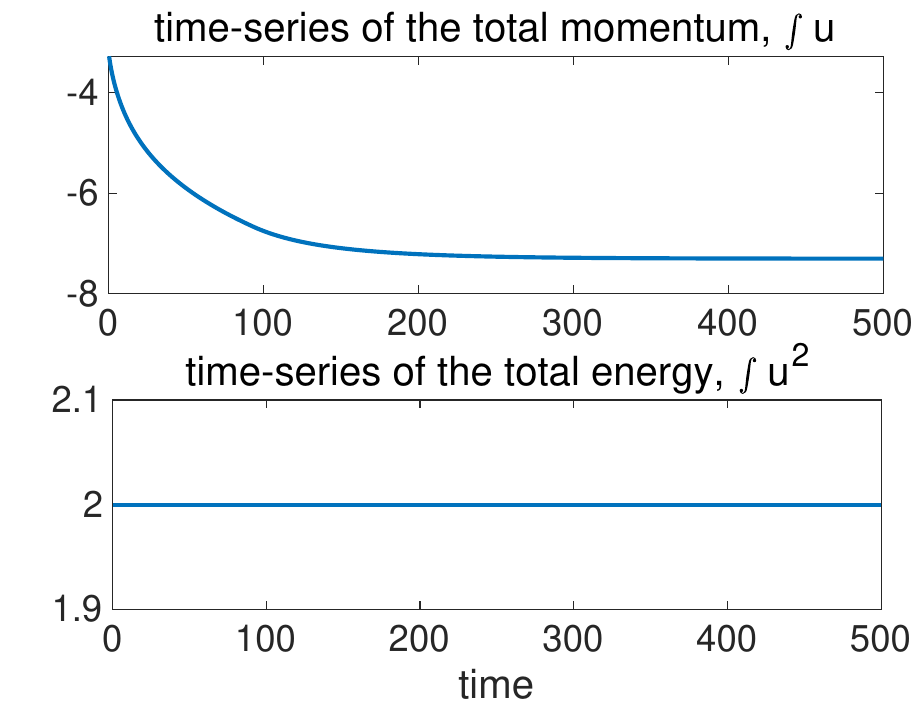}\includegraphics[scale=0.35]{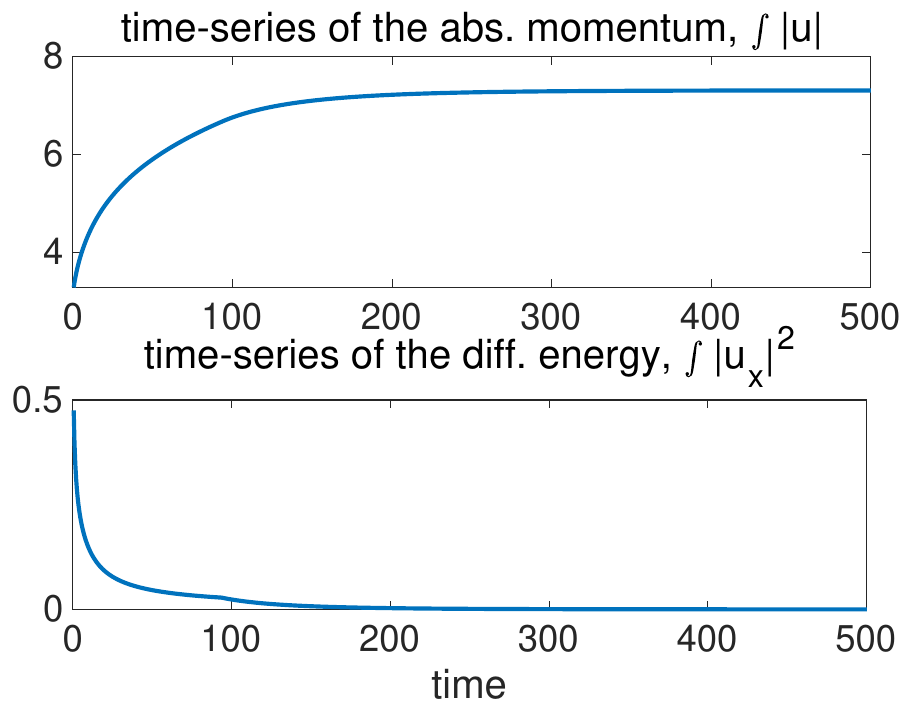}

\caption{Solution of the first-order asymptotic equation (\ref{eq:eqn_1st})
with negative initial data.\protect\label{fig:Solution-leading1}}

\end{figure}

\begin{figure}
\includegraphics[scale=0.35]{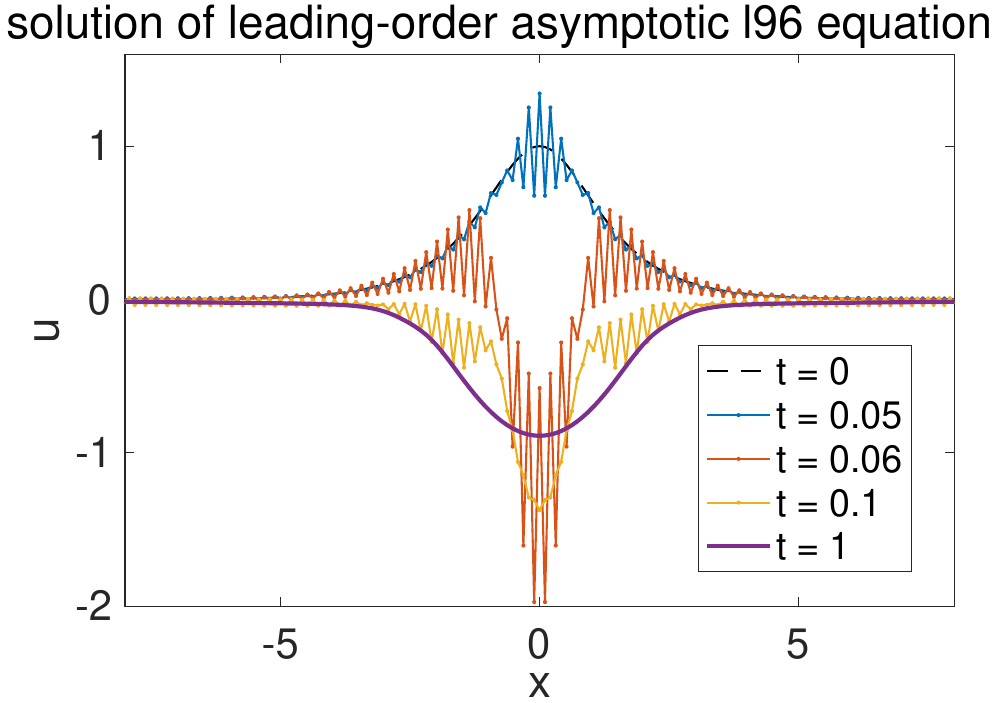}\includegraphics[scale=0.35]{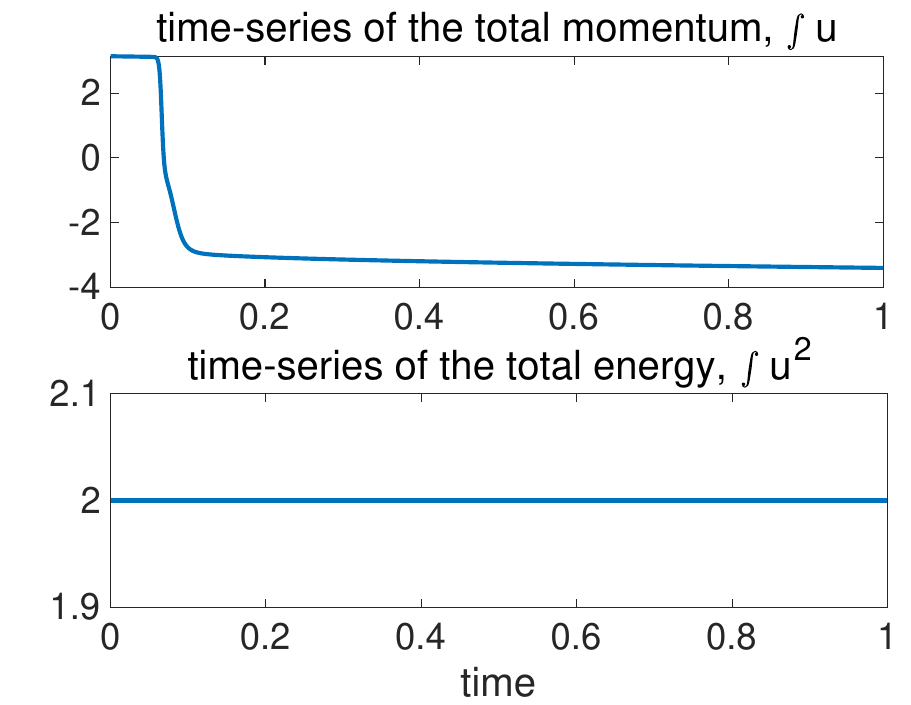}\includegraphics[scale=0.35]{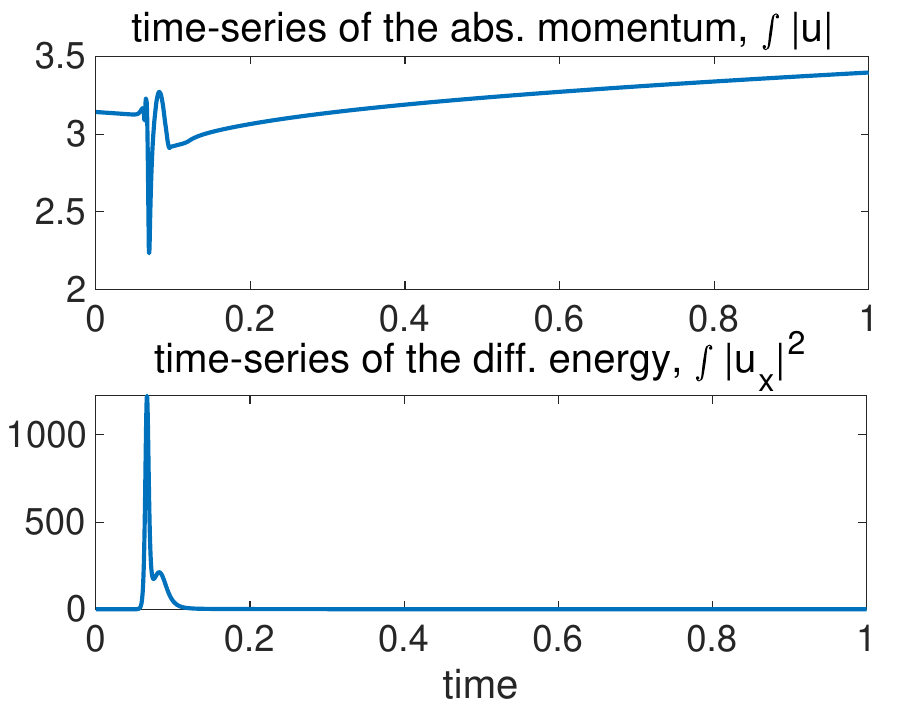}

\caption{Solution of the first-order asymptotic equation (\ref{eq:eqn_1st})
with positive initial data.\protect\label{fig:Solution-leading2}}
\end{figure}

We check our analysis results above using direct numerical simulations
of the equation (\ref{eq:eqn_1st}). A pseudo-spectral scheme with
dealiasing is used for high accuracy of the nonlinear coupling term
\cite{majda2018strategies}. We use a discretization size $J=256$.
Two different initial states with positive and negative initial values
$u_{0}\left(x\right)=\pm\mathrm{sech}\left(x\right)$ are considered.
First, Figure \ref{fig:Solution-leading1} shows the evolution of
solution starting from a negative initial state. The solution remains
in the smooth region and shows consistent performance as indicated
in Proposition \ref{prop:stable_state} and \ref{prop:negativity}
as well as the energy conservation laws \eqref{eq:conserv0} and \eqref{eq:conserv3}.
Next, the solution development from positive initial state is displayed
in Figure \ref{fig:Solution-leading2}. In this case, we look at the
transient state development from the unstable initial state. In contrast
to the negative initial value case, the transient state of the solution
demonstrates oscillations of period-two type (that is, oscillation
between adjacent grid points). This implies the downward `cascade'
of energy to the smallest scale. Finally, the small-scale oscillations
will be strongly dissipated, transferring to the final solution
in the regime with purely negative values.
\begin{rem*}
The numerical examples in Figure \ref{fig:Solution-leading1} and
\ref{fig:Solution-leading2} provide a first qualitative characterization
of `the route to chaos' in the L96 system. In the original setting
of L96 model \eqref{eq:L96} with a positive forcing $F>0$, the positive
forcing will drive the mean state $\bar{u}$ to the positive value
regime, while the first-order nonlinear coupling tend to create oscillatory
solutions during returning to the stable regime with negative values.
These competing counter effects lead to the creation of complex chaotic
feature as shown in Figure \ref{fig:Illustration-l96}.
\end{rem*}

\section{Generation of period-two oscillatory solutions\protect\label{sec:Generation-of-period-two}}

Here, we study the development of oscillatory solutions in the discrete
L96 model by considering its convergence at the continuum limit. As
observed in the numerical simulation in Figure \ref{fig:Solution-leading2},
period-two oscillating solutions in the grid size will automatically
emerge from smooth initial data and be maintained from the leading-order
nonlinear coupling effect. For the analysis, we treat the inviscid
L96 system \eqref{eq:L96-invisid} as a spatially discretized numerical
scheme in the following form
\begin{equation}
\frac{\partial u\left(x_{j},t\right)}{\partial t}=\frac{1}{ah}\left(u\left(x_{j+1},t\right)-u\left(x_{j-2},t\right)\right)u\left(x_{j-1},t\right).\label{eq:l96_num}
\end{equation}
Above, $a$ is a scaling parameter and $h=\frac{L}{J}$ is the spatial
grid size. The continuous model state $u\left(x,t\right)$ defined
on $x\in\left[0,L\right]$ is evaluated at the discretized grid points
$x_{j}=jh$ with $j=1,\cdots,J$ such that $u\left(x,t\right)$ can
be viewed as a smooth interpolation of the discrete grid values $u_{j}$.
In particular, the generalized scheme (\ref{eq:l96_num}) goes back
to the standard L96 equation (\ref{eq:L96}) by taking the parameter
values $a=3,J=40$ and $L=\frac{40}{3}$.

\subsection{Modulation equation with period-two oscillations}

We start with two local conservation laws from the semi-discrete formulation
(\ref{eq:l96_num})\addtocounter{equation}{0}\begin{subequations}\label{eq:eqn_conserv_peri2} 
\begin{align}
\frac{du_{j}^{2}}{dt}+\frac{1}{ah}\left(f_{j+\frac{1}{2}}-f_{j-\frac{1}{2}}\right) & =0,\label{eq:eqn_conserv1}\\
\frac{du_{j}}{dt}+\frac{1}{ah}\left(g_{j+\frac{1}{2}}-g_{j-\frac{1}{2}}\right) & =-\frac{h}{2a}D_{j+\frac{1}{2}}.\label{eq:eqn_conserv2}
\end{align}
\end{subequations}Above, we define the local fluxes for the energy
$u_{j}^{2}$ and the momentum $u_{j}$
\begin{equation}
f_{j+\frac{1}{2}}=-2u_{j-1}u_{j}u_{j+1},\quad g_{j+\frac{1}{2}}=-\frac{1}{2}\left(u_{j-1}u_{j}+u_{j-1}u_{j+1}+u_{j}u_{j+1}\right),\label{eq:fluxes}
\end{equation}
and the additional non-conservative term for the momentum equation
\begin{equation}
D_{j+\frac{1}{2}}=\frac{u_{j+1}-u_{j-2}}{h}\frac{u_{j}-u_{j-1}}{h}.\label{eq:damping}
\end{equation}
Notice that the local conservation equations (\ref{eq:eqn_conserv1})
and (\ref{eq:eqn_conserv2}) is consistent with the asymptotic conservation
equations \eqref{eq:conservation} in Section \ref{subsec:The-L96-system}.
The energy $u_{j}^{2}$ accepts the exact conserved form locally,
while the momentum $u_{j}$ is subject to an additional reaction term
represented by the local differences in $D_{j+\frac{1}{2}}$. As $h\rightarrow0$
and $u$ remains a classical $C^{1}$ solution, the right hand side
of (\ref{eq:eqn_conserv2}) goes to the higher-order limit, $-\frac{3h}{2a}\left(\partial_{x}u_{j}\right)^{2}$.
Thus the $C^{1}$ solution $u$ satisfies the exact conservation laws
consistent with the classical solutions of the Burgers-Hopf equation
before the formation of shocks. On the other hand, when the discontinuities
are developed, it will be shown that the damping term $D_{j+\frac{1}{2}}$
on the right hand side will give an order one contribution due to
the period-two oscillations.

To study the development of oscillations, we look at a special type
of oscillating solutions developed from the conservation equations
\eqref{eq:eqn_conserv_peri2}. We introduce the period-two states
according to two adjacent grids
\begin{equation}
v_{j+\frac{1}{2}}=\frac{1}{2}\left(u_{j}+u_{j+1}\right),\quad w_{j+\frac{1}{2}}=\frac{1}{2}\left(u_{j}^{2}+u_{j+1}^{2}\right).\label{eq:peri_2}
\end{equation}
The above two averaged states correspond to the period-two oscillating
solution referring to the alternating values between adjacent grid
points, such at $u_{j}<u_{j-1}$, $u_{j+1}>u_{j}$, and $u_{j+2}<u_{j+1}$.
Thus, we are seeking the new smooth limiting states $w,v$ that satisfy
the following period-two solution conditions
\[
u_{j+1}=v_{j+\frac{1}{2}}+\left(-1\right)^{j}\left(w_{j+\frac{1}{2}}-v_{j+\frac{1}{2}}^{2}\right)^{\frac{1}{2}},\quad M_{j+\frac{1}{2}}=\left(w_{j+\frac{1}{2}}-v_{j+\frac{1}{2}}^{2}\right)^{\frac{1}{2}}=\frac{\left(-1\right)^{j}}{2}\left(u_{j+1}-u_{j}\right).
\]
We can find the dynamical equations for the new period-two variables
$\left(w_{j+\frac{1}{2}},v_{j+\frac{1}{2}}\right)$ by adding up the
locally conserved equations (\ref{eq:eqn_conserv1}) and (\ref{eq:eqn_conserv2})
at two adjacent grid points, that is,\addtocounter{equation}{0}\begin{subequations}\label{eq:eqn_peri2} 
\begin{align}
\frac{dw_{j+\frac{1}{2}}}{dt}+\frac{1}{2ah}\left(f_{j+\frac{3}{2}}-f_{j-\frac{1}{2}}\right) & =0,\label{eq:eqn_peri2_1}\\
\frac{dv_{j+\frac{1}{2}}}{dt}+\frac{1}{2ah}\left(g_{j+\frac{3}{2}}-g_{j-\frac{1}{2}}\right) & =-\frac{h}{4a}\left(D_{j+\frac{3}{2}}+D_{j+\frac{1}{2}}\right).\label{eq:eqn_peri2_2}
\end{align}
\end{subequations}To first gain some intuition on the solutions of
the above equations, we check the steady state period-two solution,
$w_{j+\frac{1}{2}}\equiv\bar{w}$, $v_{j+\frac{1}{2}}\equiv\bar{v}$,
and $\cdots=u_{j}=u_{j+2}=\cdots<$$\cdots u_{j+1}=u_{j+3}=\cdots$
thus $M_{j+\frac{1}{2}}=u_{j+1}-u_{j}\equiv\bar{M}$. This leads to
the dynamical equations
\[
\ \frac{d\bar{v}}{dt}=\frac{1}{2ah}\bar{M}^{2}>0,\quad\mathrm{and}\quad\frac{d\bar{w}}{dt}=0.
\]
This implies how the solution performs when the period-two solution
is developed. The mean energy state $\bar{w}=2\bar{v}^{2}+\frac{1}{2}\bar{M}^{2}$
will stay as a constant while the mean amplitude of $\bar{v}$ will
increase when period-two oscillation is gradually generated in $u$.
This implies that the jump amplitude $\bar{M}$ will decrease in time,
leading to a non-oscillating solution at the final steady state.

Next, to achieve a more precise characterization for the time evolution
of the period-two oscillating solutions when discontinuity is developed
in $u$, we derive the modulation equations according to the above
semi-discrete conservation equations \eqref{eq:eqn_conserv_peri2}.
The flux terms can be expanded by Taylor series when $w,v$ are $C^{2}$
functions, that is,
\begin{align*}
-\frac{1}{2ah}\left(f_{j+\frac{3}{2}}-f_{j-\frac{1}{2}}\right) & =\frac{2}{a}\partial_{x}\left(2v^{2}-w\right)\left(v-\left(w-v^{2}\right)^{\frac{1}{2}}\right)\left(x_{j+\frac{1}{2}}\right)+O\left(h\right),\\
-\frac{1}{2ah}\left(g_{j+\frac{3}{2}}-g_{j-\frac{1}{2}}\right) & =\frac{1}{2a}\partial_{x}\left(4v^{2}-w-2v\left(w-v^{2}\right)^{\frac{1}{2}}\right)\left(x_{j+\frac{1}{2}}\right)+O\left(h\right).
\end{align*}
And the difference term on the right hand side gives
\[
-\frac{h}{4a}\left(D_{j+\frac{3}{2}}+D_{j+\frac{1}{2}}\right)=-\frac{2}{a}\partial_{x}\left(w-v^{2}\right)\left(x_{j+\frac{1}{2}}\right)+\frac{2}{a}\frac{M_{j+\frac{1}{2}}^{2}}{h}+O\left(h\right),
\]
with $M_{j+\frac{1}{2}}^{2}=\frac{1}{4}\left(u_{j+1}-u_{j}\right)^{2}=w_{j+\frac{1}{2}}-v_{j+\frac{1}{2}}^{2}\geq0$.
Notice that when the solution $u\left(x,t\right)$ is $C^{1}$, $M_{j+\frac{1}{2}}^{2}\sim\left(\partial_{x}u\right)^{2}h^{2}$,
thus the last terms on the right hand side are of the next order $O\left(h\right)$
and vanish as $h\rightarrow0$. We have the following lemma in the
region of classical solutions of $u$.
\begin{lem}
\label{lem:smooth}The solution $\left(v,w\right)$ of the modulation
state is $C^{1}$ before the formation of shock in $u$ of \eqref{eq:l96_num}.
And its leading-order equation as $h\rightarrow0$  satisfies the
following conservation form 
\begin{equation}
\begin{aligned}\partial_{t}w & =\frac{2}{a}\partial_{x}\left(2v^{3}-vw\right),\\
\partial_{t}v & =\frac{1}{a}\partial_{x}\left(2v^{2}-\frac{w}{2}\right),
\end{aligned}
\label{eq:modula_classic}
\end{equation}
where we have $u=v\pm\left(w-v^{2}\right)^{\frac{1}{2}}$.
\end{lem}

However, when the period-two oscillation is developed after the formation
of the shock, then we have $M_{j+\frac{1}{2}}^{2}\sim O\left(h\right)$.
This leads to the amplification of the period-two oscillations. In
this case, we find the \emph{modulation equations} describing the
$C^{2}$ solutions $\left(w,v\right)$ with the existence of the period-two
oscillations in $u$ at the continuum limit as $h\rightarrow0$\addtocounter{equation}{0}\begin{subequations}\label{eq:modula} 
\begin{align}
\partial_{t}w & =\frac{2}{a}\partial_{x}\left[2v^{3}-vw-\left(2v^{2}-w\right)\left(w-v^{2}\right)^{\frac{1}{2}}\right],\label{eq:modula1}\\
\partial_{t}v & =\frac{1}{a}\partial_{x}\left[4v^{2}-\frac{5w}{2}-v\left(w-v^{2}\right)^{\frac{1}{2}}\right]+\frac{S}{2a},\label{eq:modula2}
\end{align}
\end{subequations}where we define the smooth function $S\left(x,t\right)=\lim_{h\rightarrow0}\frac{\left|u_{j+1}-u_{j}\right|^{2}}{h}$
as the additional reaction term due to the period-two oscillating
solution. One first constraint for the smoothly varying states $\left(w,v\right)$
is $w\geq v^{2}$ from the definition. Above, the solution will remain
smooth when $M_{j+\frac{1}{2}}$ grows to the order $O\left(\sqrt{h}\right)$
for large amplitude period-two oscillations in $u$ solutions. We
can summarize the modulation equations \eqref{eq:modula} in the following
lemma.
\begin{lem}
\label{prop:modulation}The leading-order equation of the semi-discretized
scheme \eqref{eq:l96_num} as $h\rightarrow0$ can be written according
to the states $\mathbf{u}=\left(v,w\right)^{T}$ as
\begin{equation}
\partial_{t}\mathbf{u}-\partial_{x}\mathbf{F}=\mathbf{S},\label{eq:limit_eqn_leading}
\end{equation}
with 
\begin{equation}
\mathbf{F}=\left(f,g\right)^{T}=\frac{1}{a}\left(4v^{2}-\frac{5w}{2}-v\left(w-v^{2}\right)^{\frac{1}{2}},4v^{3}-2vw-2\left(2v^{2}-w\right)\left(w-v^{2}\right)^{\frac{1}{2}}\right)^{T}.\label{eq:conserved_flux}
\end{equation}
The additional reaction term $\mathbf{S}$ on the right hand side
satisfies:
\begin{itemize}
\item In the classical region with $u\in C^{1}$, there is no additional
source term on the right hand side $\mathbf{S=0}$;
\item In the period-two region where amplified oscillations are developed,
that is, $\left|u_{j+1}-u_{j}\right|\sim O\left(\sqrt{h}\right)$, we
have the non-zero reaction term defined as $\mathbf{S}=\frac{1}{2a}\left(\lim_{h\rightarrow0}\frac{\left|u_{j+1}-u_{j}\right|^{2}}{h},0\right)^{T}$.
\end{itemize}
\end{lem}

\begin{rem*}
1. The system \eqref{eq:limit_eqn_leading} is called strictly \emph{hyperbolic}
if the Jacobian matrix
\begin{equation}
\nabla\mathbf{F}=\begin{bmatrix}\partial_{v}f & \partial_{w}f\\
\partial_{v}g & \partial_{w}g
\end{bmatrix}=L^{-1}\Lambda L,\label{eq:hyperbolicity}
\end{equation}
is diagonalizable and has real and separated eigenvalues.

2. The additional reaction term $S$ in \eqref{eq:modula2} comes
from the discrete term $\frac{1}{h}M_{j+\frac{1}{2}}^{2}=\frac{1}{4h}\left(u_{j+1}-u_{j}\right)^{2}$.
This term will give a dominant contribution when the amplitude of
period-two oscillation $\left|u_{j+1}-u_{j}\right|$ grows large.
Still, \eqref{eq:modula} remains a good approximation to the discrete
solution with a finite grid size $h$.
\end{rem*}

\subsection{Convergence to the period-two modulation equations}

Here, we show the convergence of the discrete solution $\left(v_{j+\frac{1}{2}}\left(t\right),w_{j+\frac{1}{2}}\left(t\right)\right)$
in (\ref{eq:peri_2}) to the smooth period-two solution $\left(v\left(x_{j+\frac{1}{2}},t\right),w\left(x_{j+\frac{1}{2}},t\right)\right)$
of the modulation equations \eqref{eq:modula} at the continuum limit
as $h\rightarrow0$.

\subsubsection{Convergence with small oscillation amplitude}

First, we show that the modulation equation for $\left(v,w\right)$
gives a good approximation to the discrete period-two solution at
the small oscillation amplitude case, $\left|u_{j+1}-u_{j}\right|=O\left(\sqrt{h}\right)$.
The theorem basically follows \cite{levermore1996large} according
to the modulation equations \eqref{eq:modula}.
\begin{thm}
\label{thm:estim_peri2}Let $\mathbf{u}\left(x,t\right)=\left(v\left(x,t\right),w\left(x,t\right)\right)$
be a $C^{2}$ solution of the modulation equation \eqref{eq:modula},
and $\mathbf{U}_{j+\frac{1}{2}}=\left(v_{j+\frac{1}{2}}\left(t\right),w_{j+\frac{1}{2}}\left(t\right)\right),j=1,\cdots,J$
be the discrete period-two solution from the inviscid L96 equation
\eqref{eq:l96_num}. If both solutions belong to the hyperbolic region
(\ref{eq:hyperbolicity}) and $M_{j+\frac{1}{2}}^{2}=w_{j+\frac{1}{2}}-v_{j+\frac{1}{2}}^{2}=O\left(h\right)$
during the time interval $t\in\left[0,T\right]$, we have 
\begin{equation}
\max_{0\leq t\leq T}\left[\frac{1}{J}\sum_{j=1}^{J}\left|\mathbf{u}\left(x_{j+\frac{1}{2}},t\right)-\mathbf{U}_{j+\frac{1}{2}}\left(t\right)\right|^{2}\right]^{\frac{1}{2}}\leq C_{T}\frac{1}{J},\label{eq:peri2_bnd}
\end{equation}
where $x_{j+\frac{1}{2}}=\left(j+\frac{1}{2}\right)h$, $h=\frac{L}{J}$,
and $\left|\cdot\right|$ is the vector $L^{2}$-norm.
\end{thm}

\begin{proof}
The $C^{2}$ solution $\mathbf{u}\left(x,t\right)$ of the modulation
equation \eqref{eq:modula} can be summarized in the following equation
as
\[
\frac{d\mathbf{u}\left(x_{j+\frac{1}{2}},t\right)}{dt}=\frac{\mathbf{F}\left(\mathbf{u}\left(x_{j+\frac{3}{2}},t\right)\right)-\mathbf{F}\left(\mathbf{u}\left(x_{j-\frac{1}{2}},t\right)\right)}{2h}+\mathbf{S}_{}\left(\mathbf{u}\left(x_{j+\frac{1}{2}},t\right)\right)+O\left(h\right).
\]
Correspondingly, the solution $\mathbf{U}_{j+\frac{1}{2}}\left(t\right)$
of the discrete equations \eqref{eq:eqn_peri2} can be rewritten as
\[
\frac{d\mathbf{U}_{j+\frac{1}{2}}}{dt}=\frac{\mathbf{F}\left(\mathbf{U}_{j+\frac{3}{2}}\right)-\mathbf{F}\left(\mathbf{U}_{j-\frac{1}{2}}\right)}{2h}+\mathbf{S}\left(\mathbf{U}_{j+\frac{1}{2}}\right)+O\left(h\right).
\]
In addition, Taylor expansions from the previous computations show
that 
\[
\mathbf{F}\left(\mathbf{u}\left(x_{j+\frac{1}{2}},t\right)\right)-\mathbf{F}\left(\mathbf{U}_{j+\frac{1}{2}}\right)=\nabla\mathbf{F}_{j+\frac{1}{2}}\mathbf{e}_{j+\frac{1}{2}}+O\left(h\right),
\]
where we introduce the error $\mathbf{e}_{j+\frac{1}{2}}\left(t\right)=\mathbf{u}\left(x_{j+\frac{1}{2}},t\right)-\mathbf{U}_{j+\frac{1}{2}}\left(t\right)$
and $\nabla\mathbf{F}_{j+\frac{1}{2}}=\nabla_{\mathbf{u}}\mathbf{F}\left(\mathbf{u}\left(x_{j+\frac{1}{2}},t\right)\right)$
is the gradient about $\mathbf{u}=\left(v,w\right)$. Combining the
above equations, we have
\[
\frac{d\mathbf{e}_{j+\frac{1}{2}}}{dt}=\frac{\nabla\mathbf{F}_{j+\frac{3}{2}}\mathbf{e}_{j+\frac{3}{2}}-\nabla\mathbf{F}_{j-\frac{1}{2}}\mathbf{e}_{j-\frac{1}{2}}}{2h}+\nabla\mathbf{S}_{j+\frac{1}{2}}\mathbf{e}_{j+\frac{1}{2}}+O\left(h\right).
\]
Hyperbolicity of the equations guarantees the eigenvalue decomposition
of the coefficient matrix with real eigenvalues 
\[
\nabla\mathbf{F}_{j+\frac{1}{2}}=L_{j+\frac{1}{2}}^{-1}\Lambda_{j+\frac{1}{2}}L_{j+\frac{1}{2}}.
\]
Therefore, by introducing $\mathbf{\tilde{e}}_{j+\frac{1}{2}}=L_{j+\frac{1}{2}}\mathbf{e}_{j+\frac{1}{2}}$
we have the dynamics for the error
\begin{align*}
\frac{d\mathbf{\tilde{e}}_{j+\frac{1}{2}}}{dt}-\frac{\Lambda_{j+\frac{3}{2}}\mathbf{\tilde{e}}_{j+\frac{3}{2}}-\Lambda_{j-\frac{1}{2}}\mathbf{\tilde{e}}_{j-\frac{1}{2}}}{2h}= & L_{j+\frac{1}{2}}^{-1}\left(\frac{L_{j+\frac{3}{2}}-L_{j+\frac{1}{2}}}{2h}\Lambda_{j+\frac{3}{2}}\mathbf{\tilde{e}}_{j+\frac{3}{2}}+\frac{L_{j+\frac{1}{2}}-L_{j-\frac{1}{2}}}{2h}\Lambda_{j-\frac{1}{2}}\mathbf{\tilde{e}}_{j-\frac{1}{2}}\right)\\
 & +\frac{dL_{j+\frac{1}{2}}}{dt}L_{j+\frac{1}{2}}^{-1}\mathbf{\tilde{e}}_{j+\frac{1}{2}}+L_{j+\frac{1}{2}}\nabla\mathbf{S}_{j+\frac{1}{2}}L_{j+\frac{1}{2}}^{-1}\mathbf{\tilde{e}}_{j+\frac{1}{2}}+O\left(h\right)\\
\leq & C\left(\left|\mathbf{\tilde{e}}_{j-\frac{1}{2}}\right|+\left|\mathbf{\tilde{e}}_{j+\frac{1}{2}}\right|+\left|\mathbf{\tilde{e}}_{j+\frac{3}{2}}\right|\right)+C_{1}h.
\end{align*}
Multiplying both sides by $\tilde{\mathbf{e}}_{j+\frac{1}{2}}$ and
taking the summation about $j$ gives using the smooth dependence on $\mathbf{u}$
\[
\frac{d}{dt}\frac{1}{2}\sum_{j}\left|\mathbf{\tilde{e}}_{j+\frac{1}{2}}\right|^{2}-\sum_{j}\frac{\Lambda_{j+\frac{1}{2}}-\Lambda_{j-\frac{1}{2}}}{2h}\mathbf{\tilde{e}}_{j-\frac{1}{2}}\cdot\mathbf{\tilde{e}}_{j+\frac{1}{2}}\leq\sum_{j}\left[C\left|\tilde{\mathbf{e}}_{j+\frac{1}{2}}\right|\left(\left|\mathbf{\tilde{e}}_{j-\frac{1}{2}}\right|+\left|\mathbf{\tilde{e}}_{j+\frac{1}{2}}\right|+\left|\mathbf{\tilde{e}}_{j+\frac{3}{2}}\right|\right)+C_{1}\left|\tilde{\mathbf{e}}_{j+\frac{1}{2}}\right|h\right].
\]
 This leads to the estimate for the total error
\begin{align*}
\frac{d}{dt}\sum_{j}\left|\mathbf{\tilde{e}}_{j+\frac{1}{2}}\right|^{2} & \leq C\sum_{j}\left|\mathbf{\tilde{e}}_{j+\frac{1}{2}}\right|^{2}+hC_{1}\sum_{j}\left|\mathbf{\tilde{e}}_{j+\frac{1}{2}}\right|\\
 & \leq C^{\prime}\sum_{j}\left|\mathbf{\tilde{e}}_{j+\frac{1}{2}}\right|^{2}+C_{1}^{\prime}h.
\end{align*}
 Using Gronwall's inequality for any $t\leq T$, we have
\[
\sum_{j}\left|\mathbf{\tilde{e}}_{j+\frac{1}{2}}\right|^{2}\left(t\right)\leq C_{T}h\;\Rightarrow\;\frac{1}{J}\sum_{j}\left|\mathbf{e}_{j+\frac{1}{2}}\right|^{2}\left(t\right)\leq C_{T}h^{2}.
\]
This establishes the final result in the theorem with $C_{T}$ independent
of $h$.
\end{proof}
\begin{rem*}
In Theorem \ref{thm:estim_peri2}, we require small oscillation amplitude
$M_{j+\frac{1}{2}}\sim\left|u_{j+1}-u_{j}\right|=O\left(\sqrt{h}\right)$ in the period-two
solution while its derivative blows up $\frac{u_{j+1}-u_{j}}{h}=O\left(1/\sqrt{h}\right)$.
When the oscillations grow to larger amplitude $M_{j+\frac{1}{2}}^{2}=\frac{1}{2}\left|u_{j+1}-u_{j}\right|^{2}=O\left(1\right)$,
the unbounded $S=\frac{M^{2}}{h}$ will take over as the dominant
term and break down the period-two oscillations. Still, according
to the equation for $M$ shown next in \eqref{eq:modula3}, the time
scale of the oscillation amplitude $M$ is within $T\sim O\left(h\right)$.
We still have a bounded estimation in the error $\sum_{j}\left|\mathbf{e}_{j+\frac{1}{2}}\right|^{2}\leq\frac{C}{J}e^{TJ}\sim\frac{C^{\prime}}{J}$,
thus the solution of the modulation equation \eqref{eq:modula} can
still offer a desirable estimation to the discrete period-two solution
of \eqref{eq:l96_num} with a moderate grid size $h$ (which is in
fact the case of the standard L96 model using $J=\frac{1}{h}=40$).
\end{rem*}

\subsubsection{Approximation with large oscillation amplitude}

Next, we can check the development of large period-two oscillations
by introducing an additional equation for $M=\left(w-v^{2}\right)^{\frac{1}{2}}$.
When strong oscillations are induced, we assume $M_{j+\frac{1}{2}}=\frac{\left(-1\right)^{j}}{2}\left(u_{j+1}-u_{j}\right)$
describing the jump amplitude between two adjacent grids. The dynamical
equation for the difference can be written down as
\[
\frac{dM_{j+\frac{1}{2}}}{dt}=\frac{\left(-1\right)^{j}}{2ah}\left[\left(u_{j+2}-u_{j-1}\right)u_{j}-\left(u_{j+1}-u_{j-2}\right)u_{j-1}\right].
\]
Through Taylor expansion of the difference form up to $O\left(h\right)$,
the continuum equation is reached as 
\begin{equation}
\partial_{t}M=-\frac{2}{ah}vM-\frac{1}{2a}\partial_{x}\left(M^{2}-v^{2}+w\right).\label{eq:modula3}
\end{equation}
The first term $-\frac{2v}{ah}M$ on the right hand side of \eqref{eq:modula3}
shows that the period-two oscillation will be amplified in the domain
where $v<0$, while the oscillations will be damped where $v>0$.
In addition, the advection term implies that the period-two oscillations
will propagate with velocity $v$. Especially, combining with \eqref{eq:modula},
we can rewrite the closed system for $\left(v,M\right)$ with finite
oscillation amplitude
\begin{equation}
\begin{aligned}\partial_{t}v & =\frac{2M^{2}}{ah}+\frac{1}{a}\partial_{x}\left(\frac{3}{2}v^{2}+\frac{5}{2}M^{2}-vM\right),\\
\partial_{t}M & =-\frac{2v}{ah}M-\frac{1}{a}\partial_{x}M.
\end{aligned}
\label{eq:closed_modula}
\end{equation}
Above, we assume $M\sim O\left(1\right)$ grows to an order one term,
and the equations for $v$ and $M$ have the stiff leading order term
dependent on the grid size $h^{-1}$. 

Then, we can separate the leading-order term and investigate its effect
on the particular solution $\left(v_{h},M_{h}\right)$
\begin{equation}
\partial_{t}v_{h}=\frac{2M_{h}^{2}}{ah},\quad\partial_{t}M_{h}=-\frac{2v_{h}}{ah}M_{h}.\label{eq:osci_leading}
\end{equation}
Directly integrating the above equation leads to the following lemma
about a closed form of the solution of \eqref{eq:osci_leading}.
\begin{lem}
\label{lem:explicit}The leading-order equations \eqref{eq:osci_leading}
for $\left(v_{h},M_{h}\right)$ have the explicitly integrable solution
\begin{equation}
v_{h}\left(x,t\right)=\lambda\left(x\right)\frac{1-\mu\left(x\right)\exp\left(-\frac{4\lambda}{ah}t\right)}{1+\mu\left(x\right)\exp\left(-\frac{4\lambda}{ah}t\right)},\quad M_{h}\left(x,t\right)=\frac{2\lambda\left(x\right)\sqrt{\mu\left(x\right)}}{1+\mu\left(x\right)\exp\left(-\frac{4\lambda}{ah}t\right)}\exp\left(-\frac{2\lambda}{ah}t\right),\label{eq:soln_leading}
\end{equation}
where $\lambda,\mu\geq0$ are smooth functions determined by the initial
values of $v_{h},M_{h}$. Further in the full equations \eqref{eq:closed_modula},
if $v=v_{h}+O\left(h\right)$, there is also $M=M_{h}+O\left(h\right)$.
\end{lem}

\begin{proof}
Equations \eqref{eq:osci_leading} are actually a coupled ODE system
and satisfy
\[
\partial_{t}^{2}v_{}=-\frac{2}{ah}\partial_{t}v_{}^{2}\;\Rightarrow\;\partial_{t}v_{}=-\frac{2}{ah}v_{}^{2}+C\left(x\right).
\]
The constant can be found from the initial value $C=\partial_{t}v\left(x,0\right)+\frac{2}{ah}v^{2}\left(x,0\right)>0$,
with a small enough $h$. For convenience, we introduce $C=\frac{2\lambda^{2}}{ah}$.
Therefore, by integrating the above equation again about $v$ and
$t$, we find
\[
\left|\frac{\lambda+v}{\lambda-v}\right|=\left|\frac{\lambda+v_{0}}{\lambda-v_{0}}\right|e^{\frac{2}{ah}t}=\mu^{-1}e^{\frac{4\lambda}{ah}t}\;\Rightarrow\;v=\lambda\frac{1-\mu e^{-\frac{4\lambda}{ah}t}}{1+\mu e^{-\frac{4\lambda}{ah}t}}.
\]
And the expression for $M$ follows immediately from the above solution
of $v$.

Next, let $v=v_{h}+\tilde{v}h$ and $M=M_{h}+\tilde{M}h$. In the
equation for $M$ in \eqref{eq:closed_modula}, the last advection
term can be canceled by changing $M\left(x,t\right)\rightarrow M\left(x+\frac{1}{a}t,t\right)$.
Thus, the equation satisfies using the solution \eqref{eq:osci_leading}
\[
\partial_{t}\tilde{M}=-\frac{2}{ah}\left(v_{h}+\tilde{v}h\right)\tilde{M}-\frac{2}{ah}M_{h}\tilde{v}.
\]
Integrating the equation in time, we get
\begin{align*}
\left|\tilde{M}\left(t\right)\right| & \leq\frac{2}{ah}\int_{0}^{t}e^{-\frac{2}{ah}\int_{s}^{t}\left(v_{h}\left(\tau\right)+\tilde{v}\left(\tau\right)h\right)d\tau}M_{h}\left(s\right)\left|\tilde{v}\left(s\right)\right|ds\\
 & \leq\frac{2}{ah}\int_{0}^{t}e^{\frac{\lambda}{ah}s+C}M_{h}\left(s\right)\left|\tilde{v}\left(s\right)\right|ds\\
 & \leq\frac{C}{h}\int_{0}^{t}\exp\left(-\frac{\lambda}{ah}s\right)ds\leq C^{\prime}.
\end{align*}
Above, the second line uses the explicit solution of $v_{h}$, $\left|\int_{s}^{t}v_{h}\left(\tau\right)d\tau\right|\leq\frac{\lambda}{2}s+C_{1}h$,
and the uniform boundedness of $\tilde{v}$, $\left|\int_{s}^{t}\tilde{v}\left(\tau\right)hd\tau\right|\leq C_{2}h$.
And the third line again uses the explicit solution of $M_{h}$ and
uniform bound of $\tilde{v}$. Thus we show $\tilde{M}$ is also uniformly
bounded from $h$.
\end{proof}
Lemma \ref{lem:explicit} provides a more precise estimate for the
development of oscillation amplitude $M$ dependent on the grid size
$h$. This implies that the leading-order equation gives the uniformly
bounded solution $v_{h}$ and the uniformly decaying solution $M_{h}$
independent of the stiff factor $h^{-1}$. According to this leading-order
performance, we can introduce the new modulation equations for $\left(v_{},w_{}\right)$
dependent on finite grid size $h$
\begin{equation}
\begin{aligned}\partial_{t}w_{} & =\frac{2}{a}\partial_{x}\left[v\left(2v^{2}-w\right)-\left(2v_{}^{2}-w_{}\right)\left(w_{}-v_{}^{2}\right)^{\frac{1}{2}}\right],\\
\partial_{t}v_{} & =\frac{2\left(w_{}-v_{}^{2}\right)}{ah}+\frac{1}{a}\partial_{x}\left[4v_{}^{2}-\frac{5w_{}}{2}-v\left(w_{}-v_{}^{2}\right)^{\frac{1}{2}}\right].
\end{aligned}
\label{eq:modula4}
\end{equation}
With this, we can generalize the result in Theorem \ref{thm:estim_peri2}
to the time with large period-two oscillation amplitude, $\left|u_{j+1}-u_{j}\right|=O\left(1\right)$.
\begin{cor}
\label{cor:converg_high}Let $\mathbf{u}_{}=\left(v_{},w_{}\right),t\in\left[0,T\right]$
be the $C^{2}$ solution of the modulation equation \eqref{eq:modula4},
and $\mathbf{U}_{j+\frac{1}{2}}=\left(v_{j+\frac{1}{2}},w_{j+\frac{1}{2}}\right),j=1,\cdots,J$
the discrete period-two solution from the inviscid L96 equation \eqref{eq:l96_num}.
Assume that $v$ and $w$ are both uniformly bounded for any $h>0$
and $v=v_{h}+O\left(h\right)$ for any $0\leq t\leq T$, then we have
\begin{equation}
\max_{0\leq t\leq T}\left[\frac{1}{J}\sum_{j=1}^{J}\left|\mathbf{u}\left(x_{j+\frac{1}{2}},t\right)-\mathbf{U}_{j+\frac{1}{2}}\left(t\right)\right|^{2}\right]^{\frac{1}{2}}\leq C_{T}\frac{1}{J}.\label{eq:peri2_bnd-1}
\end{equation}
\end{cor}

\begin{proof}
Notice that the only difference in the new modulation equations \eqref{eq:modula4}
is the new factor $\frac{1}{h}$ in the reaction equation. Thus the
equation for the error $\mathbf{e}_{j+\frac{1}{2}}\left(t\right)=\mathbf{u}\left(x_{j+\frac{1}{2}},t\right)-\mathbf{U}_{j+\frac{1}{2}}\left(t\right)$
becomes
\[
\frac{d\mathbf{e}_{j+\frac{1}{2}}}{dt}=\frac{1}{h}M_{j+\frac{1}{2}}\nabla\mathbf{M}_{j+\frac{1}{2}}\mathbf{e}_{j+\frac{1}{2}}+\frac{\nabla\mathbf{F}_{j+\frac{3}{2}}\mathbf{e}_{j+\frac{3}{2}}-\nabla\mathbf{F}_{j-\frac{1}{2}}\mathbf{e}_{j-\frac{1}{2}}}{2h}+O\left(h\right),
\]
where $\mathbf{M}_{j+\frac{1}{2}}=\frac{2}{a}\left(M,0\right)^{T}\left(x_{j+\frac{1}{2}},t\right)$
and $M^{2}=w-v^{2}$. Using the explicit leading-order solution $M_{h}$
and Lemma \ref{lem:explicit}, we have $\left|M_{j+\frac{1}{2}}\right|\leq C\exp\left(-\frac{2\bar{\lambda}}{ah}t\right)+Ch$
and $\bar{\lambda}=\min_{j}\lambda_{j+\frac{1}{2}}$. Following the
same line of argument as in Theorem \ref{thm:estim_peri2}, we get
the error estimate 
\[
\frac{d\mathbf{e}_{j+\frac{1}{2}}}{dt}\leq\left[C_{1}+\frac{C_{}}{h}\exp\left(-\frac{2\bar{\lambda}}{ah}t\right)\right]\left(\left|\mathbf{e}_{j-\frac{1}{2}}\right|+\left|\mathbf{e}_{j+\frac{1}{2}}\right|+\left|\mathbf{e}_{j+\frac{3}{2}}\right|\right)+C_{2}h.
\]
Finally, Gronwall's inequality yields
\[
\sum_{j}\left|\mathbf{e}_{j+\frac{1}{2}}\right|^{2}\left(t\right)\leq\left[C_{T}+\exp\left(\frac{C}{h}\int_{0}^{T}e^{-\frac{2\lambda}{ah}t}dt\right)\right]h\leq\left[C_{T}+\exp\left(\frac{a}{2\bar{\lambda}}\right)\right]h=C_{T}^{\prime}h.
\]
\end{proof}
\begin{rem*}
It is found in the numerical simulations that it is usually the driven
effect of the reaction term $\frac{M^{2}}{h}$ rather than the moving
out of the hyperbolic region that breaks down the period-two oscillatory
solutions. The leading $O\left(h^{-1}\right)$ terms combined with
the additional $O\left(1\right)$ terms on the right hand sides of
\eqref{eq:closed_modula} may lead to complicated coupling dynamics,
thus finally drive the solution away from the small perturbation region
$v=v_{h}+O\left(h\right)$ required in the above theorem. Then, the
clean period-two solution will break down to create fully chaotic
features.
\end{rem*}

\subsection{Persistence and breakdown of period-two oscillations}

Now if we consider the leading flux term $\mathbf{F}$ in the modulation
equation \eqref{eq:conserved_flux}, the Jacobian matrix \eqref{eq:hyperbolicity}
becomes
\[
\nabla\mathbf{F}=\frac{1}{a}\begin{bmatrix}8v-\left(w-v^{2}\right)^{\frac{1}{2}}+v^{2}\left(w-v^{2}\right)^{-\frac{1}{2}} & -\frac{5}{2}-\frac{1}{2}v\left(w-v^{2}\right)^{-\frac{1}{2}}\\
12v^{2}-2w-8v\left(w-v^{2}\right)^{\frac{1}{2}}+2v\left(2v^{2}-w\right)\left(w-v^{2}\right)^{-\frac{1}{2}} & -2v+2\left(w-v^{2}\right)^{\frac{1}{2}}-\left(2v^{2}-w\right)\left(w-v^{2}\right)^{-\frac{1}{2}}
\end{bmatrix}.
\]
Solving the eigenvalues of the above matrix reveals that the condition
for hyperbolicity of the modulation equation is always satisfied when
$w-v^{2}>0$, thus the solution always remains in the hyperbolic region.
Still, if we consider the weakly oscillatory region up to order $O\left(\sqrt{h}\right)$
and neglect the higher-order terms due to $w-v^{2}=O\left(h\right)$,
the hyperbolic region with real eigenvalues requires
\begin{equation}
w^{2}>4v\left(3w-2v^{2}\right)\left(w-v^{2}\right)^{\frac{1}{2}}.\label{eq:hyper_leading}
\end{equation}
This gives the condition for maintaining only weakly period-two oscillations
within $\left|u_{j+1}-u_{j}\right|=O\left(\sqrt{h}\right)$. The hyperbolic
region and the development of oscillatory solutions are illustrated
in Figure \ref{fig:Illustration-hyper}. The period-two oscillations
gradually developed large amplitudes and moved out of the weak oscillation
hyperbolic region. This leads to a dominant reaction term $\frac{M^{2}}{h}$
that finally destroyed the period-two oscillation.

To see the breakdown of the period-two oscillations more clearly,
it can be found directly from \eqref{eq:modula3} that
\[
\frac{1}{2}\partial_{t}M^{2}=-\frac{2v}{ah}M^{2}-\frac{1}{2a}\partial_{x}M^{2}\;\Rightarrow\;\frac{d}{dt}\frac{1}{2}\int_{\Omega}M^{2}dx=-\frac{2}{ah}\int_{\Omega}vM^{2}dx,
\]
where the integration is taken in the oscillatory region $\Omega_{}$
so that $M^{2}\mid_{\partial\Omega}=0$ . Thus a necessary condition
of generating growing oscillation amplitude $M$ requires $v<0$ at
some point of the domain. In addition, it implies that the period-two
oscillation will emerge at an approximate growth rate of $\left|\frac{v}{h}\right|$
depending on the discrete grid size. As the amplitude of oscillations
grows, $v$ will finally reach positive values at some point. Then,
the periodic-two oscillation will starts to get damped at the points
where $v$ reaches positive values. These features are explicitly
observed in the numerical experiments in Figure \ref{fig:States-period2}
and are consistent with the leading-order solution \eqref{eq:soln_leading}.

In addition, we can check the instability in the semi-discrete system
(\ref{eq:l96_num}) around a steady mean state $\bar{u}$. As in the
previous section, introduce the mean-fluctuation decomposition of
the state $u\left(x_{j},t\right)=\bar{u}+\tilde{u}_{j}\left(t\right)$.
The linearized equation of \eqref{eq:l96_num} becomes
\[
\frac{d\tilde{u}_{j}}{dt}=\frac{1}{ah}\bar{u}\left(\tilde{u}_{j+1}-\tilde{u}_{j-2}\right).
\]
Assuming the plane wave solution $\tilde{u}_{j}=e^{i\left(kx_{j}-\omega_{k}t\right)}$,
we find the dispersion relation
\begin{align*}
\omega_{k} & =\frac{\bar{u}}{ah}\left[-\left(\sin kh+\sin2kh\right)+i\left(\cos kh-\cos2kh\right)\right]\\
 & =\frac{\bar{u}}{ah}\left[-\sin kh\left(1+2\cos kh\right)+i\left(-2\cos^{2}kh+\cos kh+1\right)\right].
\end{align*}
Positive growth rate for the fluctuation modes is induced if the imaginary
part $\mathrm{Im}\omega_{k}>0$. Thus we have that instability is
induced when $\bar{u}>0,\cos kh>-\frac{1}{2}$ or $\bar{u}<0,\cos kh<-\frac{1}{2}$.
Especially, in the case when $\bar{u}<0$, the maximum growth rate
$c_{*}=\frac{-2\bar{u}}{ah}$ is reached at the critical wavenumber
$k_{*}=\frac{\pi}{h}$ and we have $\mathrm{Re}\omega_{k_{*}}=0$.
The corresponding critical plane wave solution becomes
\[
\tilde{u}_{j}^{*}=e^{i\left(k_{*}jh-\omega_{k_{*}}t\right)}=e^{\frac{-2\bar{u}}{ah}t}\left(-1\right)^{j}.
\]
This implies that the period-two oscillation can be excited automatically
from a negative mean velocity $\bar{u}<0$. This is also confirmed
in the numerical results in Figure \ref{fig:States-period2} where
the period-two oscillations are amplified in the region with $\bar{u}<0$.

To summarize, we can describe the evolution of the solution of (\ref{eq:l96_num})
in the following three stages, as the classical region, period-two
region, and finally the fully chaotic region:
\begin{itemize}
\item \emph{Stage I. Smooth state $u$ in the starting time}: the state
$u\left(x,t\right)$ remains as a $C^{2}$ function, so the solution
performs as the Burgers-Hopf equation until the development of discontinuity;
\item \emph{Stage II. Development of period-two oscillation solution}: small
amplitude period-two oscillations are developed in $u$ at the point
of discontinuity, while the modulation states $\left(v,w\right)$
stay as $C^{2}$ functions;
\item \emph{Stage III. Breakdown of period-two solution}: Period-two oscillations
increase to the positive region with $v>0$ and finally get damped.
The states $\left(v,w\right)$ break down from the period-two oscillations,
thus fully chaotic solution begins to develop. 
\end{itemize}

\subsection{Numerical verification for the development of oscillatory solutions}

\begin{figure}
\subfloat{\includegraphics[scale=0.34]{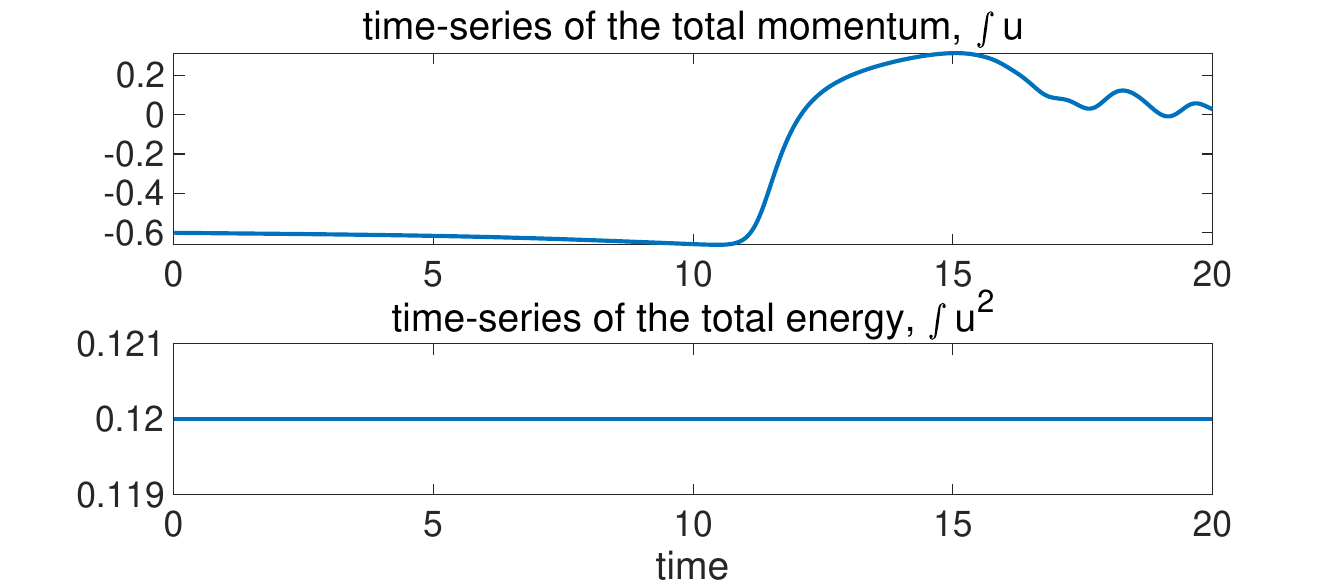}\includegraphics[scale=0.34]{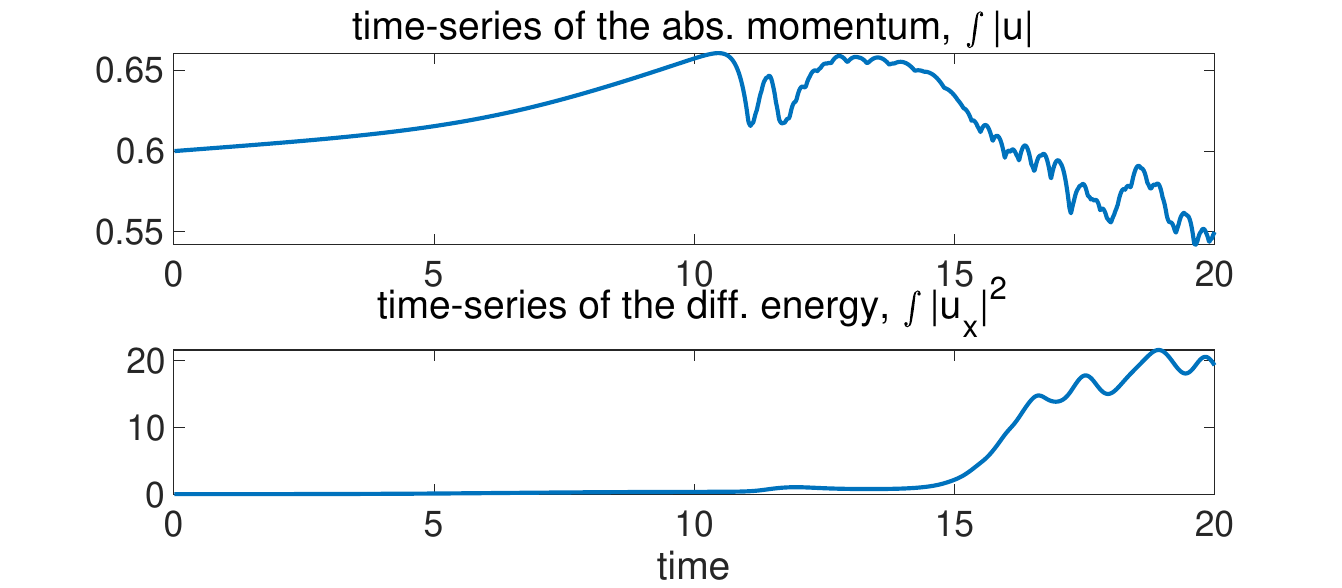}}

\subfloat{\includegraphics[scale=0.35]{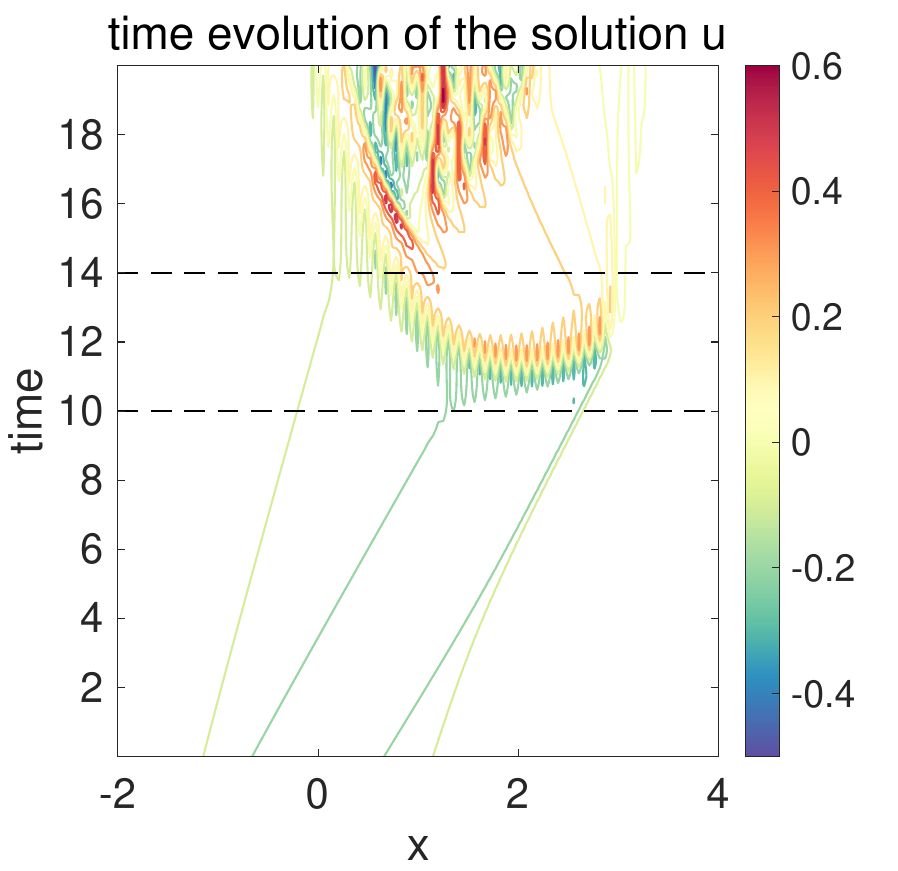}\includegraphics[scale=0.35]{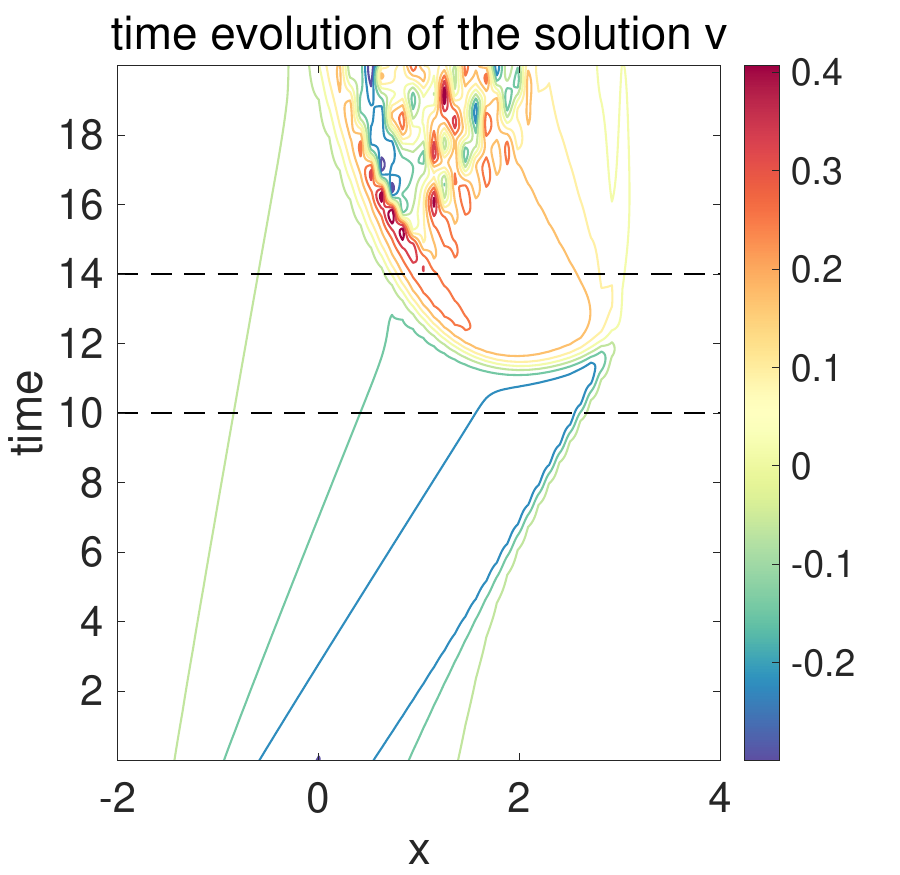}\includegraphics[scale=0.35]{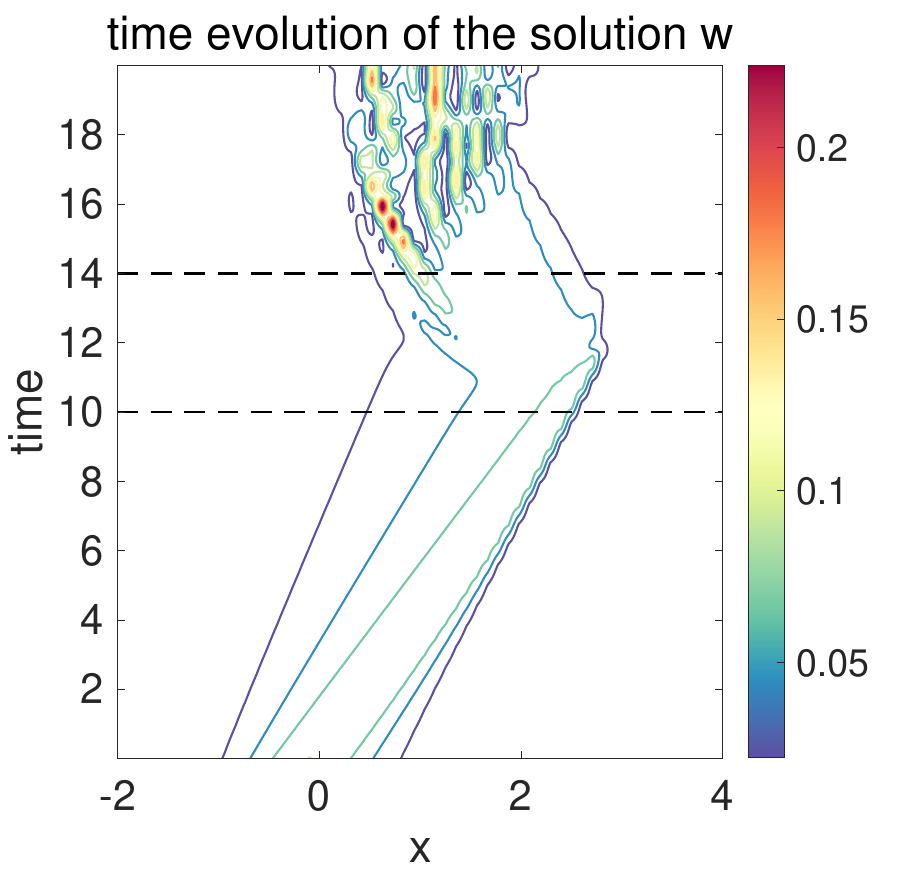}}

\caption{Time-series of the key model integrals and the evolution of the period-two
solution from smooth initial state in the inviscid L96 model \eqref{eq:l96_num}.\protect\label{fig:Time-evolution-1layer}}

\end{figure}

In the numerical experiments, we run the L96 model \eqref{eq:l96_num}
with discretization $J=256$ and model parameters $a=3$ and $h=8/J$.
The initial state is taken as $u_{0}\left(x\right)=-0.3\mathrm{sech}^{2}\left(\frac{x}{2}\right)$.
We pick the negative initial value since the period-two solution can
be induced from $\bar{u}<0$ from the linear instability. The 4th-order
Runge-Kutta method is used for the time integration with time step
$\Delta t=1\times10^{-3}$ to achieve desirable accuracy. First, the
time evolution of total momentum $\int udx$ and total energy $\int u^{2}dx$
as well as the related quantities are plotted in the upper panel of
Figure \ref{fig:Time-evolution-1layer}. The solution starts with
the classical solution with smooth $u$. Consistent with our analysis
in Section \ref{sec:Leading-order-equations}, the total energy is
strictly conserved, while the total momentum is slowly decaying due
to the damping from $-\int u_{x}^{2}dx$. $\int\left|u\right|dx$
is also increasing in the classical region since we have $u<0$ in
the initial time. 

\begin{figure}
\includegraphics[scale=0.38]{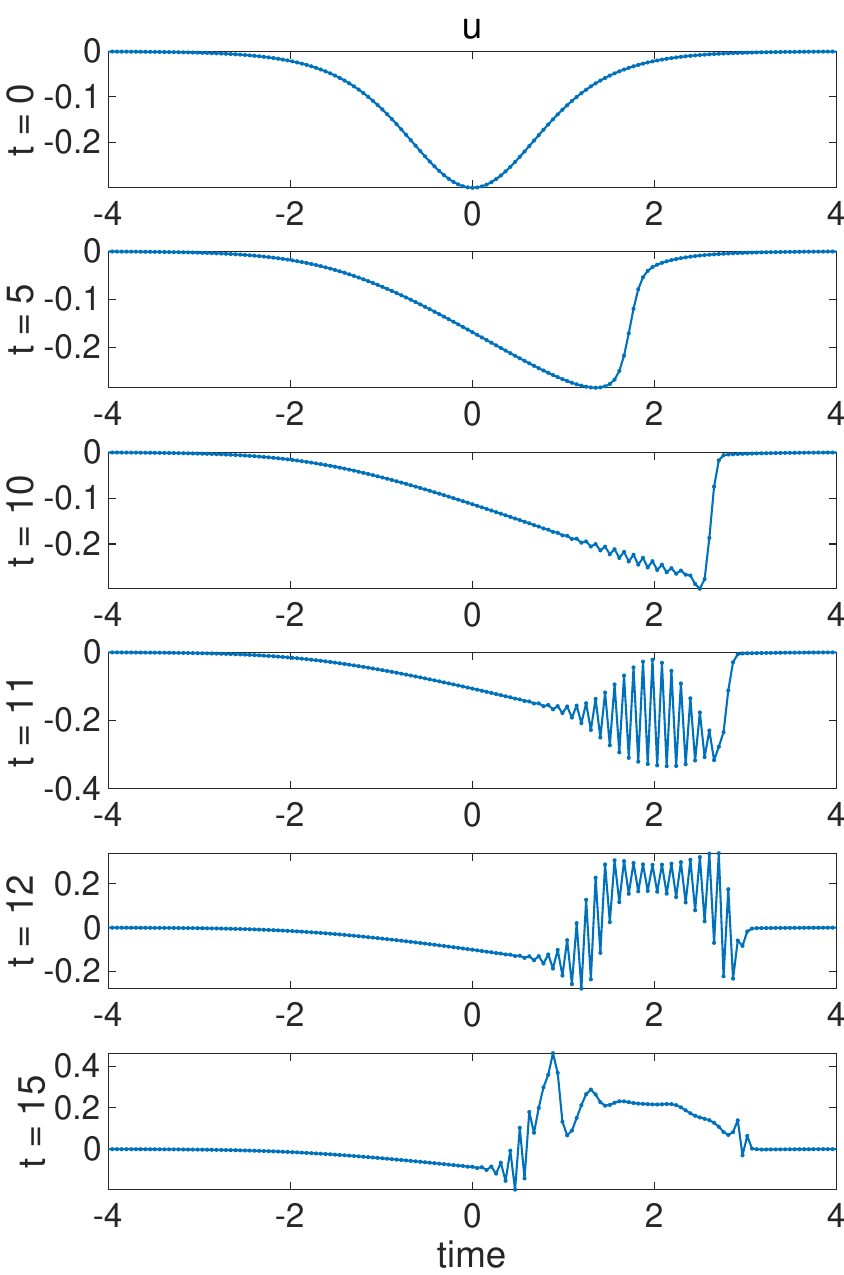}\includegraphics[scale=0.38]{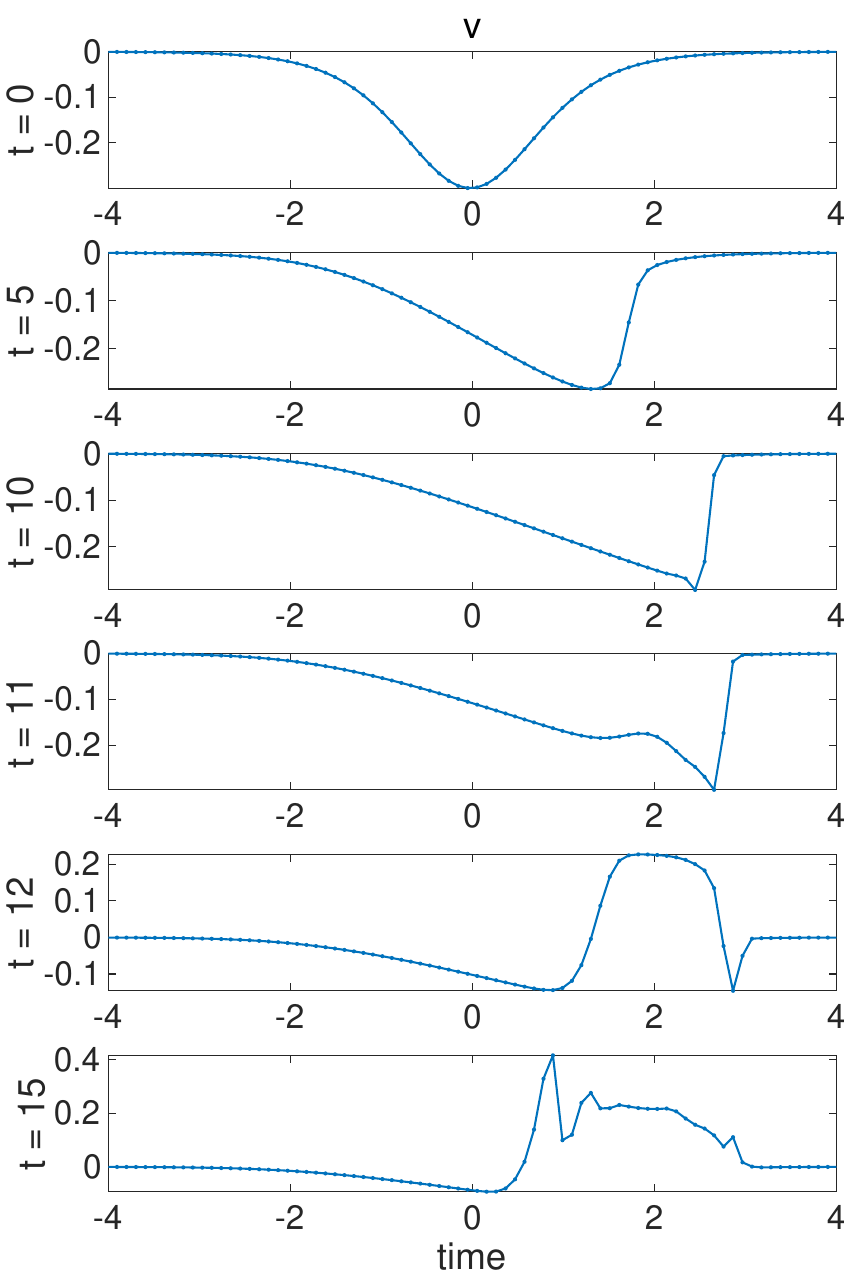}\includegraphics[scale=0.38]{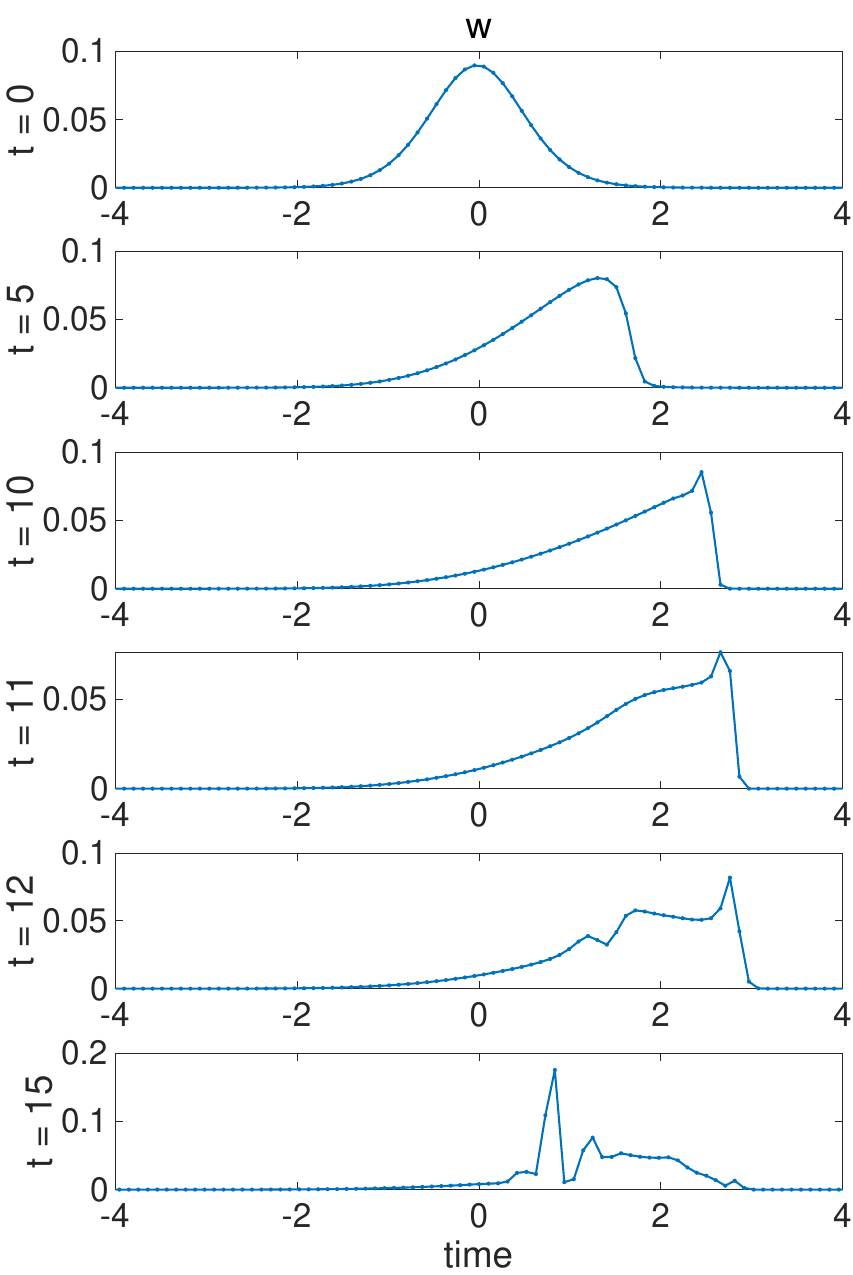}

\caption{States $u,v,w$ at several different time instants during the development
of break down of period-two solution.\protect\label{fig:States-period2}}
\end{figure}

Next, a shock is developed in $u$ at around $t=10$. This leads to
the oscillations in $u$ and the creation of period-two solution.
This can be observed in the lower panel of Figure \ref{fig:Time-evolution-1layer}
for the time evolutions of $u$ and $v,w$ and more clearly in the
several time snapshots of $u,v,w$ in Figure \ref{fig:States-period2}.
Especially, we observe that $v$ starts to increase in this region
due to the reaction term $S>0$ in the equation for of $v$. Period-two
oscillations at the grid size are automatically developed at the left
side of the discontinuity and quickly get amplified. At the same time,
the period-two solutions $v,w$ remain smooth except at the point
of discontinuity.

\begin{figure}
\includegraphics[scale=0.3]{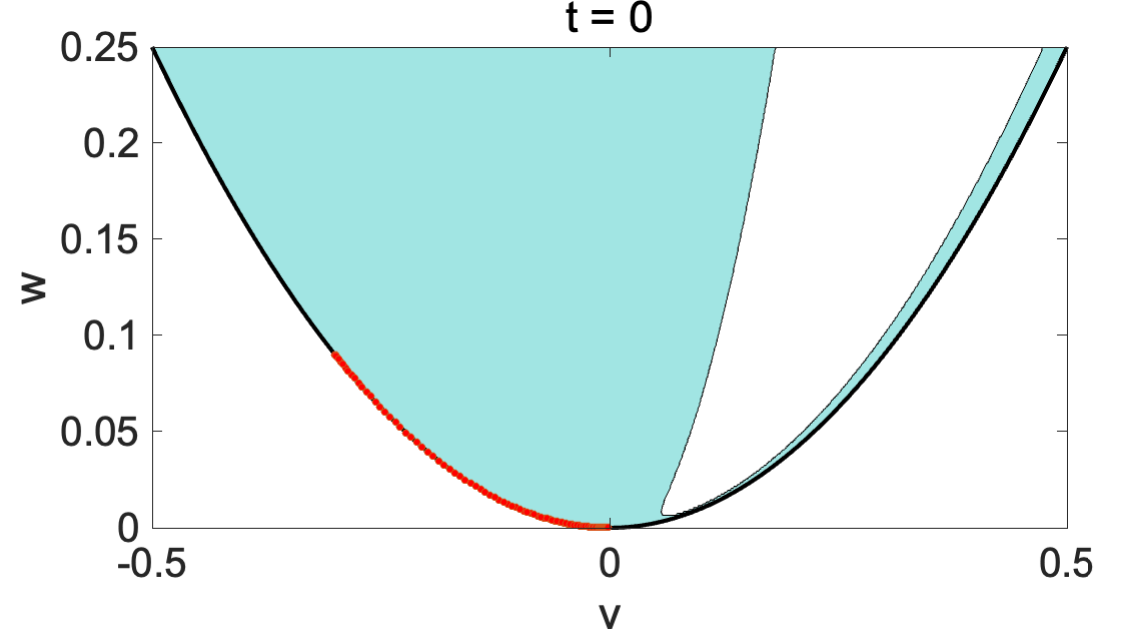}\includegraphics[scale=0.3]{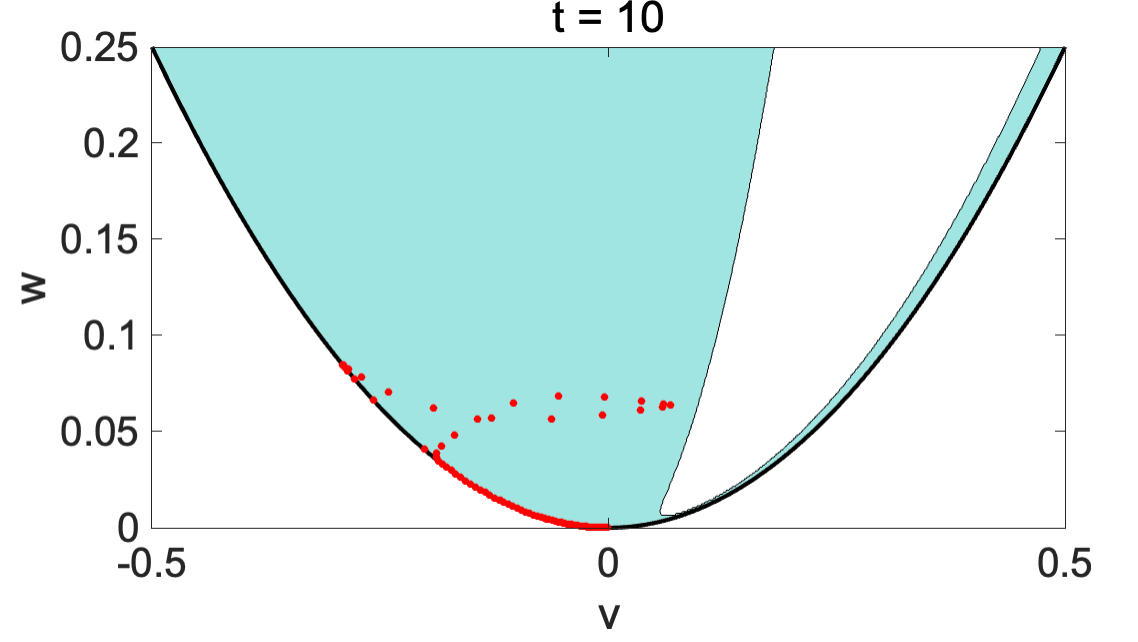}\includegraphics[scale=0.3]{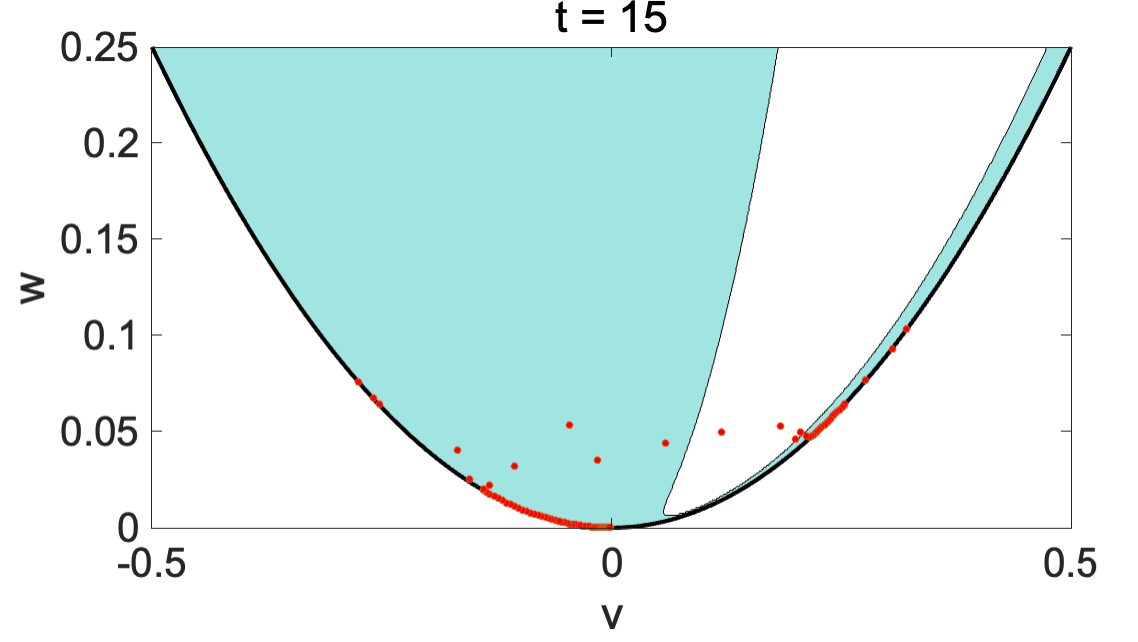}

\caption{The hyperbolic region \eqref{eq:hyper_leading} in shaded color and
black solid line shows $w=v^{2}$. Several typical solutions $\left(v,w\right)$
are also plotting illustrating the development of period-two solutions.\protect\label{fig:Illustration-hyper}}

\end{figure}

Finally, $u$ evolves into positive values with the increasing oscillation
amplitudes, and the period-two oscillations in the region with $v>0$
get damped. The smooth period-two solutions $v,w$ break down and
fully chaotic behaviors of the solution start to emerge. This indicates
the generation of many complex features as observed in the L96 system.
We further show in Figure \ref{fig:Illustration-hyper} the evolution
of solutions $\left(v,w\right)$ beyond the hyperbolic region computed
in \eqref{eq:hyper_leading}. It shows that the oscillatory period-two
solutions gradually develop in amplitude and evolve out to the fully
chaotic behavior. 

\section{The two-layer Lorenz 96 system and period-three oscillations\protect\label{sec:The-two-layer-Lorenz}}

\begin{figure}
\includegraphics[scale=0.42]{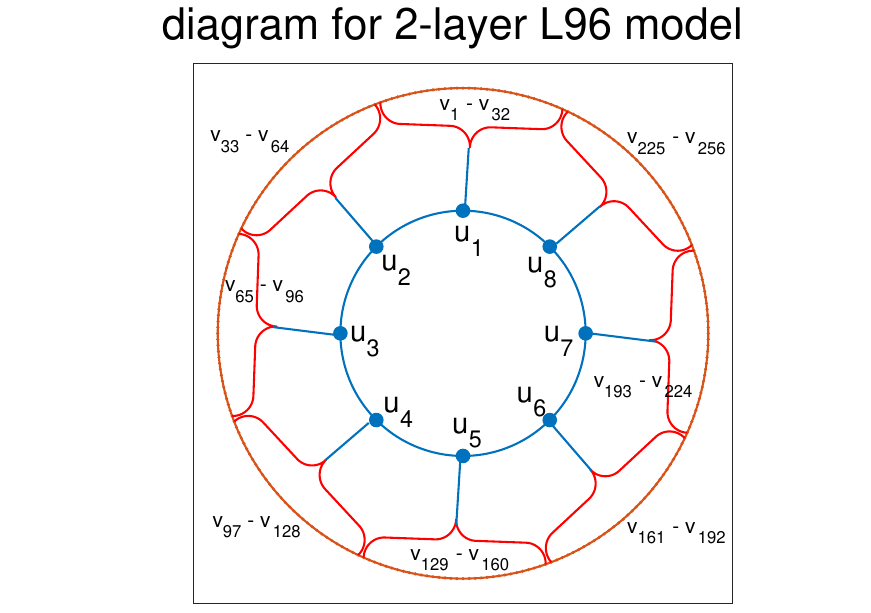}\hspace{-2em}\includegraphics[scale=0.42]{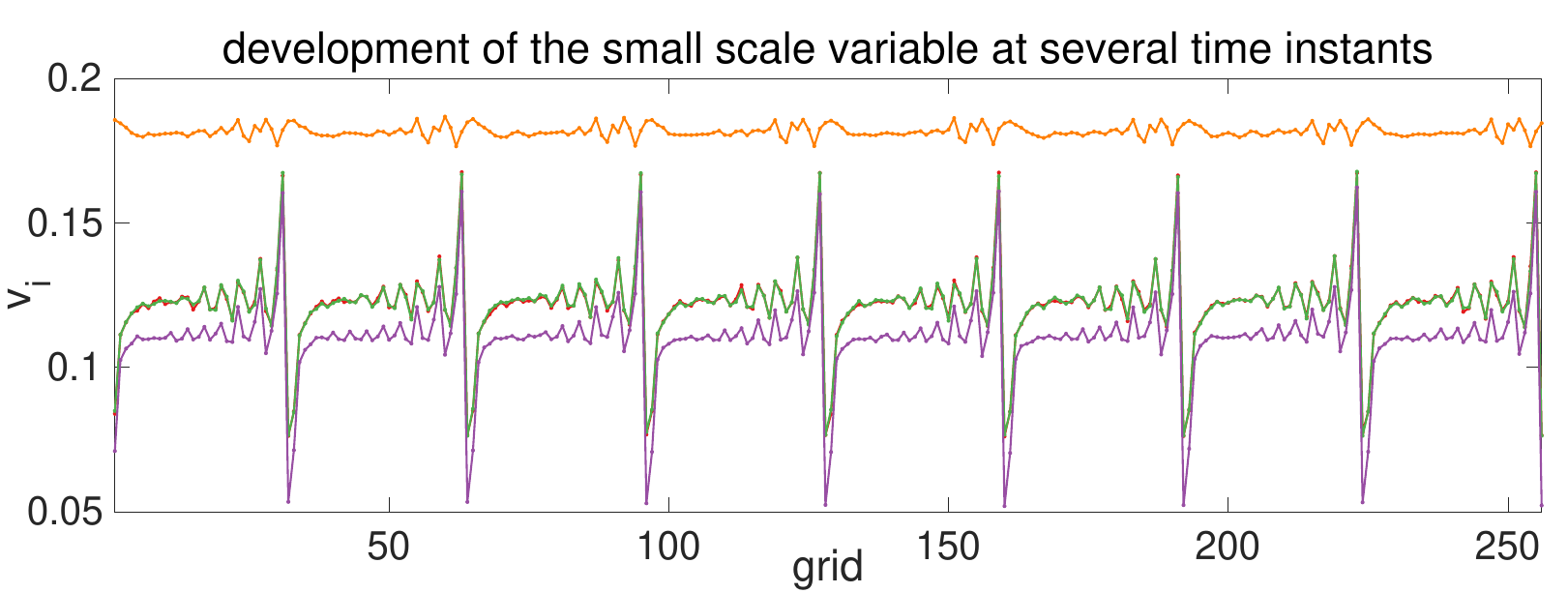}

\caption{Diagram to illustrate the coupling structure of the 2-layer L96 model
(\ref{eq:2layer}) and the solution of the small-scale state $v_{i}$
at several time with $J=8,L=32$.\protect\label{fig:Diagram-to-illustrate}}

\end{figure}

As a further generalization of the original one-layer system, the
two-layer L96 system \cite{arnold2013stochastic,wilks2005effects}
introduces an additional second layer $v_{j,l}$ to each of the original
L96 system state $u_{j},j=1,\cdots,J$ such that 
\begin{equation}
\begin{aligned}\frac{du_{j}}{dt} & =\left(u_{j+1}-u_{j-2}\right)u_{j-1}-du_{j}+F-\frac{\tilde{h}\tilde{c}}{\tilde{b}}\sum_{s=1}^{L}v_{j,s}\\
\frac{dv_{j,l}}{dt} & =-\tilde{c}\tilde{b}\left(v_{j,l+2}-v_{j,l-1}\right)v_{j,l+1}-\tilde{c}v_{j,l}+\frac{\tilde{h}\tilde{c}}{\tilde{b}}u_{j}.
\end{aligned}
\label{eq:2layer}
\end{equation}
The new layer variables $v_{j,l}=v_{i}$ with $l=1,\cdots,L$ can
be viewed as small scales with the `stretched' reorganized index $i=l+L\left(j-1\right)$.
Periodic boundary conditions are used for both the two sets of variables,
$u_{j+J}=u_{j}$ and $v_{i+JL}=v_{i}$. In general, $u_{j}$ states
are large-amplitude and low-frequency, each of which is coupled to
a branch of the small-amplitude high-frequency variables $v_{j,l}$.
Notice that the second layer states $v_{j,l}$ in the two-layer L96
model above are only locally coupled with the corresponding first
layer state $u_{j}$. In the model parameters, $\tilde{h}$ is the
coupling coefficient, $\tilde{b}$ is the spatial-scale ratio, and
$\tilde{c}$ is the time-scale ratio. The large and small scale coupling
structure is illustrated in Figure \ref{fig:Diagram-to-illustrate}.

A typical solution of the small-scale state $v_{i}$ at several time
instants is illustrated in Figure \ref{fig:Diagram-to-illustrate}.
It is observed that oscillatory solutions are generated at the boundaries
of the large-scale state $u_{j}$ similar to that in the one-layer
model. However, it also shows that the oscillations are no longer
within the grid size, while in particular it appears that the period-three
solution will emerge in this case due to the large and small scale
coupling of states.

\subsection{Strong scale separation with two interacting large-scale states}

In analyzing oscillatory solutions of the two-layer L96 model as it
approaches the continuous limit, we again focus on the nonlinear and
multiscale coupling terms (that is, neglecting the forcing and damping
effects in both large and small scales of \eqref{eq:2layer}), and
introduce the new parameters indicating the two different scales explicitly
as
\begin{equation}
\begin{aligned}\frac{du\left(X_{j},t\right)}{dt} & =\left(u\left(X_{j+1},t\right)-u\left(X_{j-2},t\right)\right)u\left(X_{j-1},t\right)-h\sum_{l=1}^{L}v\left(X_{j},x_{l},t\right)\\
\frac{dv\left(X_{j},x_{l},t\right)}{dt} & =-\frac{\gamma}{h}\left(v\left(X_{j},x_{l+2},t\right)-v\left(X_{j},x_{l-1},t\right)\right)v\left(X_{j},x_{l+1},t\right)+\gamma u\left(X_{j},t\right).
\end{aligned}
\label{eq:l96-multi}
\end{equation}
Above, we introduce the large-scale coordinate $X=\epsilon x$ with
$\epsilon$ indicating the scale separation and the small-scale resolution
$h$. In the new model parameters, we have the large-scale discretization
$\Delta x=\frac{D}{J}$ and small-scale discretization $h=\frac{\Delta x}{L}=\frac{D}{JL}$
(next we take the domain size $D=1$ for simplicity). Thus, we have
the large-scale state $u=u\left(X,t\right)$ and small-scale state
$v=v\left(X,x,t\right)$. By scaling the previous discrete equations
(\ref{eq:2layer}) with the rescaled variables $u_{j}\rightarrow u\left(X_{j}\right)$
and $v_{j,l}\rightarrow\epsilon^{\frac{1}{2}}v\left(X_{j},x_{l}\right)$,
we find the relation between the old and new parameters as
\[
\epsilon=\left(\tilde{c}\tilde{b}\right)^{-2},\quad\frac{h}{L}=\tilde{h}\tilde{b}^{-2},\quad\epsilon=\left(\tilde{c}^{2}\tilde{h}\right)^{-1}h=\left(\tilde{c}^{2}\tilde{h}\right)^{-1}\frac{h}{L},\quad\gamma=\tilde{c}^{2}\tilde{h}.
\]
It can be found that the multiscale equations \eqref{eq:l96-multi}
still maintain the conservation of total energy
\[
\frac{dE}{dt}=0,\quad\mathrm{with}\quad E=\frac{\gamma}{2}\sum_{j}u_{j}^{2}+\frac{1}{2}\sum_{j,l}v_{j,l}^{2}.
\]

Here for simplicity, we assume that there are only two states $\left(u_{1},u_{2}\right)$
in the large scale. Accordingly, the small scale state $v\left(x,t\right)$
can be decomposed into two regimes, with $v_{1,l}$ associated with
$u_{1}$ and $v_{2,l}$ associated with $u_{2}$. First, we look at
the large-scale motion of the states. Define
\begin{equation}
\begin{aligned}v_{1,l} & =\bar{v}_{1}+v_{1,l}^{\prime},\quad\bar{v}_{1}=\frac{1}{L}\sum_{l=1}^{L}v_{1,l},\\
v_{2,l} & =\bar{v}_{2}+v_{2,l}^{\prime},\quad\bar{v}_{2}=\frac{1}{L}\sum_{l=1}^{L}v_{2,l}.
\end{aligned}
\label{eq:scale-2layer}
\end{equation}
Substituting the above decomposition \eqref{eq:scale-2layer} into
\eqref{eq:l96-multi}, we get the coupling equation for the two large-scale
states $u_{1},u_{2}$
\begin{equation}
\frac{du_{1}}{dt}=\left(u_{2}^{2}-u_{1}u_{2}\right)-\frac{1}{2}\bar{v}_{1},\quad\frac{du_{2}}{dt}=\left(u_{1}^{2}-u_{1}u_{2}\right)-\frac{1}{2}\bar{v}_{2},\label{eq:mean-large}
\end{equation}
where $\bar{v}_{1},\bar{v}_{2}$ give the upscale feedback to the
large-scale states.

In particular with a strong scale separation, by letting $L\rightarrow\infty$
and $h\rightarrow0$, the small-scale state $v$ goes to the continuum
limit, denoted as $v\left(x,t\right)=v_{1},x<0$ and $v\left(x,t\right)=v_{2},x<0$.
This leads to the same large-scale equations \eqref{eq:mean-large}
with the up-scale feedback as the continuum limit of \eqref{eq:scale-2layer}
consistent with the discrete case in \eqref{eq:scale-2layer}
\begin{equation}
\bar{v}_{1}=\frac{2}{D}\int_{-D/2}^{0}v\left(x,t\right)dx,\quad\bar{v}_{2}=\frac{2}{D}\int_{0}^{D/2}v\left(x,t\right)dx.\label{eq:l96_u}
\end{equation}
Correspondingly, the small-scale state $v\left(x,t\right)$ is defined
on the periodic domain $\left[-\frac{D}{2},\frac{D}{2}\right]$ and
follows the continuous equation
\begin{equation}
\partial_{t}v+3\gamma v\partial_{x}v=\gamma f_{u}-\frac{3}{2}\left(v\partial_{xx}v+2\left(\partial_{x}v\right)^{2}\right)\gamma^{2}\epsilon-\left[\frac{3}{2}\left(v\partial_{xxx}v+2\partial_{x}v\partial_{xx}v\right)\gamma^{3}\epsilon^{2}+O\left(\epsilon^{3}\right)\right],\label{eq:l96_v}
\end{equation}
where we introduce $f_{u}=u_{1}$ when $x<0$ and $f_{u}=u_{2}$ when
$x>0$, and the scaling parameter $\gamma=\epsilon^{-1}h$. The high-order
terms on the right hand side of the above equation will vanish as
$\epsilon\rightarrow0$. Through direct computation, we find the conservation
of the total energy in the above coupled equations \eqref{eq:mean-large}
and \eqref{eq:l96_v}
\begin{equation}
\frac{d}{dt}\left[\frac{\gamma}{2}\left(u_{1}^{2}+u_{2}^{2}\right)+\frac{1}{2}\int_{-D/2}^{D/2}v^{2}dx\right]=0,\label{eq:total_ene}
\end{equation}
together with the detailed up and down scale coupling dynamics
\[
\frac{d}{dt}\gamma\left(\frac{u_{1}^{2}+u_{2}^{2}}{2}\right)=-\frac{d}{dt}\frac{1}{2}\int_{-D/2}^{D/2}v^{2}dx=-\frac{u_{1}\bar{v}_{1}+u_{2}\bar{v}_{2}}{2}.
\]
Notice that the second-order term for the small-scale equation keeps
exactly same form as in the one-layer continuous equation. Thus, discussions
for conservation laws can be inherited here. 

\begin{figure}
\includegraphics[scale=0.3]{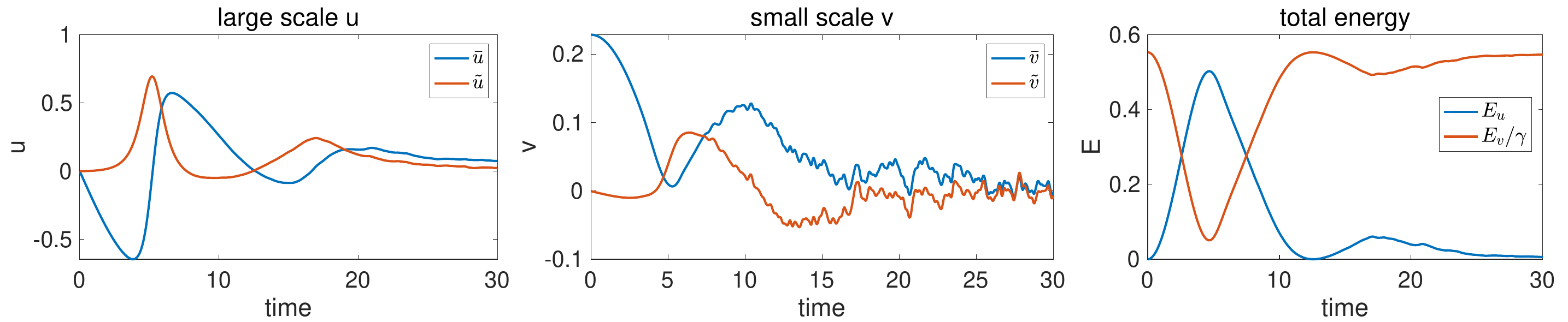}

\includegraphics[scale=0.3]{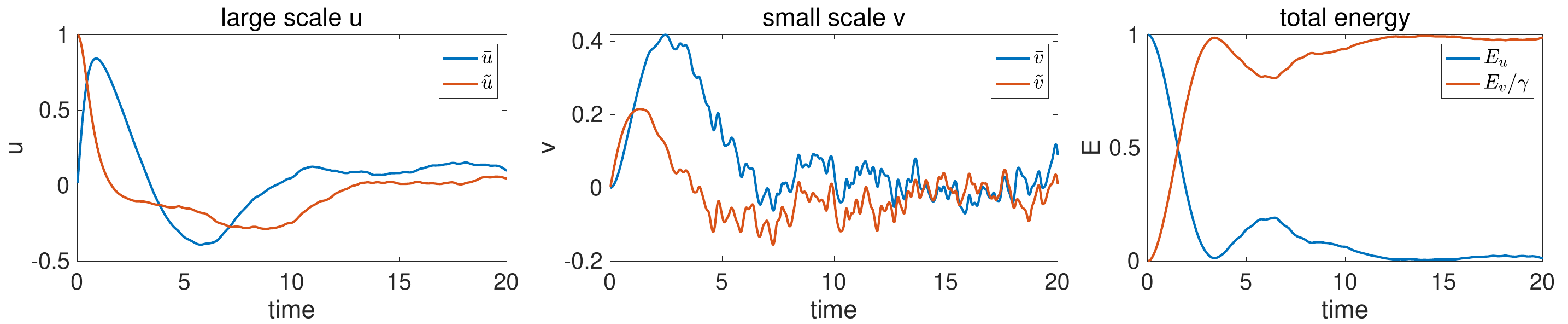}

\caption{Time evolution of the large-scale state $\bar{u}=\frac{1}{2}\left(u_{1}+u_{2}\right)$,
$\tilde{u}=\frac{1}{2}\left(u_{1}-u_{2}\right)$ and the small-scale
state $\bar{v}=\frac{1}{2}\left(\bar{v}_{1}+\bar{v}_{2}\right)$,
$\tilde{v}=\frac{1}{2}\left(\bar{v}_{1}-\bar{v}_{2}\right)$ as well
as the corresponding energy in large and small scales in two test
cases.\protect\label{fig:Time-evolution-2layer}}
\end{figure}

In addition, we can derive the upscaling equations for the slow-varying
large-scale states $\bar{v}=\frac{1}{2}\left(\bar{v}_{1}+\bar{v}_{2}\right)$,
$\tilde{v}=\frac{1}{2}\left(\bar{v}_{1}-\bar{v}_{2}\right)$ and $\bar{u}=\frac{1}{2}\left(u_{1}+u_{2}\right)$,
$\tilde{u}=\frac{1}{2}\left(u_{1}-u_{2}\right)$ at the leading-order
as
\begin{equation}
\begin{aligned}\frac{d\bar{v}}{dt} & =\gamma\bar{u},\;\frac{d\tilde{v}}{dt}=\gamma\tilde{u}-3\gamma\left(v_{0}^{2}-v_{1}^{2}\right),\\
\frac{d\bar{u}}{dt} & =-\bar{v}+2\tilde{u}^{2},\quad\frac{d\tilde{u}}{dt}=-\tilde{v}-2\tilde{u}\bar{u}.
\end{aligned}
\label{eq:large_contium}
\end{equation}
where $v_{0}\left(t\right)=v\left(0,t\right)$ and $v_{1}\left(t\right)=v\left(1/2,t\right)$
are the boundary fluxes from the small-scale feedback. Through this
way of strong scale separation, we are able to decompose the small-scale
state $v_{l}=\bar{v}+v_{l}^{\prime}$, and the residual fluctuation
modes will return to the similar one-layer equation. Figure \ref{fig:Time-evolution-2layer}
first illustrates the competing effects of the large and averaged
small-scale states in two typical test cases described in Section
\ref{subsec:Numerical-verification}. It demonstrates the typical
energy cascades from $u$ to $v$ ending up in the fully chaotic region
with most energy in small scales $v$.

\subsection{Period-three modulation equations in small scales from weakly nonlinear
analysis}

Next, we consider the evolution of small-scale fluctuations. According
to the large-scale coupling in \eqref{eq:large_contium}, the large
scale states $u_{1},u_{2}$ act as the forcing effect creating a discontinuity
in $v$ at the boundary $x=0$. Thus, we can again study the small-scale
dynamics separately based on the local decomposition $v_{l}=\bar{v}+v_{l}^{\prime}$
around a steady state $\bar{v}$
\begin{equation}
\frac{dv_{l}^{\prime}}{dt}=-\frac{\gamma}{h}\left(\bar{v}+v_{l+1}^{\prime}\right)\left(v_{l+2}^{\prime}-v_{l-1}^{\prime}\right).\label{eq:fluc-eqn}
\end{equation}
Let's start with the linearized equation of \eqref{eq:fluc-eqn} around
a constant $\bar{v}$
\[
\frac{dv_{l}^{\prime}}{dt}=-\frac{\gamma}{h}\bar{v}\left(v_{l+2}^{\prime}-v_{l-1}^{\prime}\right),
\]
and seek the plain wave solution of the form $v_{l}^{\prime}=\exp\left(i\left(kx_{l}-\omega_{k}t\right)\right)$
with $x_{l}=lh$. By directly substituting the plain wave solution
in the above linearized equation, we find the dispersion relation
\[
\omega_{k}=-i\frac{\gamma\bar{v}}{h}e^{-ikh}\left(e^{i3kh}-1\right),\quad c_{k}=\frac{d\omega_{k}}{dk}=\gamma\bar{v}e^{-ikh}\left(2e^{i3kh}+1\right).
\]
This implies a first steady state solution corresponding to $k_{0}=0$,
$\omega_{0}=0$ and $c_{0}=3\gamma\bar{v}$. In addition, two other
steady state plain wave solutions will emerge, with $k_{1}=\frac{2\pi}{3h}$,
$\omega_{1}=0,c_{1}=3\gamma\bar{v}e^{-i\frac{2\pi}{3}}$ and $k_{2}=\frac{4\pi}{3h}$,
$\omega_{2}=0,c_{2}=3\gamma\bar{v}e^{i\frac{2\pi}{3}}$. Therefore,
$\omega_{1}$ and $\omega_{2}$ represent the existence of two periodic-three
solutions $v^{1},v^{2}$ with a constant group velocity together with
the constant mode $v^{0}$
\[
v_{l}^{0}=1,\quad v_{l}^{1}=e^{i\frac{2\pi}{3}l},\quad v_{l}^{2}=e^{i\frac{4\pi}{3}l}.
\]

According to the steady state solutions, we can perform weakly nonlinear
analysis around the steady mean state $\bar{v}$. Assume that the
fluctuation state $v_{l}^{\prime}$ consists of a uniform state $\eta$
together with the two coupling period-three solutions $\xi,\zeta$,
that is,
\begin{equation}
v_{l}^{\prime}\left(t\right)=\eta_{l}\left(t\right)+e^{i\frac{2\pi}{3}l}\xi_{l}\left(t\right)+e^{i\frac{4\pi}{3}l}\zeta_{l}\left(t\right).\label{eq:peri3-soln}
\end{equation}
We consider the real solution $v_{l}^{\prime}$ so only taking the
real part of the base functions in the above solution. Thus, putting
\eqref{eq:peri3-soln} back into the fluctuation equation \eqref{eq:fluc-eqn}
and combining the common terms with the same frequency, we find the
equations for $\eta,\xi,\zeta$ as an envelope of the rapidly oscillating
fluctuations 
\begin{equation}
\begin{aligned}\frac{d\eta_{l}}{dt}= & -\gamma\left(\bar{v}+\eta_{l+1}\right)\frac{\eta_{l+2}-\eta_{l-1}}{h}+\frac{\gamma}{2}\xi_{l+1}\frac{\zeta_{l+2}-\zeta_{l-1}}{h}+\frac{\gamma}{2}\zeta_{l+1}\frac{\xi_{l+2}-\xi_{l-1}}{h},\\
\frac{d\xi_{l}}{dt}= & \frac{\gamma}{2}\left(\bar{v}+\eta_{l+1}\right)\frac{\xi_{l+2}-\xi_{l-1}}{h}-\gamma\zeta_{l+1}\frac{\zeta_{l+2}-\zeta_{l-1}}{h}+\frac{\gamma}{2}\xi_{l+1}\frac{\eta_{l+2}-\eta_{l-1}}{h},\\
\frac{d\zeta_{l}}{dt}= & \frac{\gamma}{2}\left(\bar{v}+\eta_{l+1}\right)\frac{\zeta_{l+2}-\zeta_{l-1}}{h}-\gamma\xi_{l+1}\frac{\xi_{l+2}-\xi_{l-1}}{h}+\frac{\gamma}{2}\zeta_{l+1}\frac{\eta_{l+2}-\eta_{l-1}}{h}.
\end{aligned}
\label{eq:eqn-peri3}
\end{equation}
Using \eqref{eq:eqn-peri3}, we study the evolution of small fluctuations
localized around a constant steady state $\bar{v}$ in the small-scale
variable $v$. Therefore with a bit abuse of notation, we introduce
the following smooth functions $\eta\left(z,\tau\right),\xi\left(z,\tau\right),\zeta\left(z,\tau\right)$
as the continuum limit of the three discrete fluctuation modes $\eta_{l},\xi_{l},\zeta_{l}$
\begin{equation}
\eta_{l}=h^{2}\eta\left(x_{l}+c_{}t,h^{2}t\right),\;\xi_{l}=h\xi\left(x_{l}+c_{}t,h^{2}t\right),\;\zeta_{l}=h\zeta\left(x_{l}+c_{}t,h^{2}t\right).\label{eq:scales-fluc}
\end{equation}
Above, we focus on the coupling between the uniform mode $\eta$ with
the two period-three modes $\xi,\zeta$, and we introduce the slowly
moving frame with a constant velocity $-c$.

Then, the governing equations for the three modes can be discovered
through asymptotic expansions of each order terms in \eqref{eq:eqn-peri3}.
First, the leading-order $O\left(1\right)$ terms give
\[
\left(c+3\gamma\bar{v}\right)\partial_{z}\eta=\frac{3}{2}\gamma\partial_{z}\left(\xi\zeta\right),\quad c\partial_{z}\xi=\frac{3}{2}\gamma\bar{v}\partial_{z}\xi,\quad c\partial_{z}\zeta=\frac{3}{2}\gamma\bar{v}\partial_{z}\zeta.
\]
Above, $\xi$ and $\zeta$ always satisfy the same equation, thus
we only need to keep one of them. The second identity above tells
the moving frame velocity $c$. By integrating about $z$ and using
the vanishing boundary condition, the leading order solution gives
\begin{equation}
c=\frac{3}{2}\gamma\bar{v},\quad\eta=\frac{1}{3\bar{v}}\xi\zeta.\label{eq:asymp_O0}
\end{equation}
Second, the $O\left(h\right)$ terms give
\[
\frac{\bar{v}}{2}\partial_{z}^{2}\xi=\partial_{z}\left(\zeta^{2}\right),\quad\frac{\bar{v}}{2}\partial_{z}^{2}\zeta=\partial_{z}\left(\xi^{2}\right),
\]
Again by integrating the first two identities about $z$ and using
the relation \eqref{eq:asymp_O0}, we get the relations between the
two period-three solutions
\begin{equation}
\xi^{2}=\frac{\bar{v}}{2}\partial_{z}\zeta,\;\zeta^{2}=\frac{\bar{v}}{2}\partial_{z}\xi.\label{eq:asymp_O1}
\end{equation}
At last, the $O\left(h^{2}\right)$ terms give
\[
\begin{aligned}\partial_{\tau}\xi= & \frac{3}{4}\gamma\bar{v}\partial_{z}^{3}\xi+\frac{3}{2}\gamma\partial_{z}\left(\eta\xi\right)-\frac{3}{2}\gamma\left[\left(\partial_{z}\zeta\right)^{2}+\partial_{z}\left(\zeta\partial_{z}\zeta\right)\right],\\
\partial_{\tau}\zeta= & \frac{3}{4}\gamma\bar{v}\partial_{z}^{3}\zeta+\frac{3}{2}\gamma\partial_{z}\left(\eta\zeta\right)-\frac{3}{2}\gamma\left[\left(\partial_{z}\xi\right)^{2}+\partial_{z}\left(\xi\partial_{z}\xi\right)\right].
\end{aligned}
\]
Using the identities in \eqref{eq:asymp_O0} and \eqref{eq:asymp_O1},
it yields the final coupled equations for $\xi,\zeta$
\begin{equation}
\partial_{\tau}\xi=-\frac{11\gamma}{\bar{v}^{2}}\xi^{4}-\frac{5\gamma}{\bar{v}}\zeta\xi\partial_{z}\xi,\quad\partial_{\tau}\zeta=-\frac{11\gamma}{\bar{v}^{2}}\zeta^{4}-\frac{5\gamma}{\bar{v}}\xi\zeta\partial_{z}\zeta.\label{eq:asymp_O2}
\end{equation}
Furthermore, if we define $\rho=\xi+\zeta$ and $\theta=\xi\zeta$,
we can derive the following equivalent equations using \eqref{eq:asymp_O2}
and \eqref{eq:asymp_O1} as the modulation equation for the period-three
oscillations
\begin{equation}
\begin{aligned}\partial_{\tau}\rho= & -\frac{5\gamma}{\bar{v}}\theta\partial_{z}\rho-\frac{11\gamma}{4}\left(\partial_{z}\rho\right)^{2}+\frac{22\gamma}{\bar{v}^{2}}\theta^{2},\\
\partial_{\tau}\theta= & -\frac{5\gamma}{\bar{v}}\theta\partial_{z}\theta-\frac{11\gamma}{2\bar{v}}\theta\rho\partial_{z}\rho+\frac{11\gamma}{\bar{v}^{2}}\rho\theta.
\end{aligned}
\label{eq:asymp-coupled}
\end{equation}

Therefore, the asymptotic approximate solution of the small-scale
variable in \eqref{eq:fluc-eqn} can be written as
\begin{equation}
\begin{aligned}v_{l}= & \bar{v}+\cos\left(\frac{2\pi}{3}l\right)\left[\xi\left(x_{l}+\frac{3}{2}\gamma\bar{v}t,h^{2}t\right)+\zeta\left(x_{l}+\frac{3}{2}\gamma\bar{v}t,h^{2}t\right)\right]h\\
+ & \frac{1}{3\bar{v}}\xi\left(x_{l}+\frac{3}{2}\gamma\bar{v}t,h^{2}t\right)\zeta\left(x_{l}+\frac{3}{2}\gamma\bar{v}t,h^{2}t\right)h^{2}+O\left(h^{3}\right)\\
= & \bar{v}+\cos\left(\frac{2\pi}{3}l\right)\rho\left(x_{l}+\frac{3}{2}\gamma\bar{v}t,h^{2}t\right)h+\frac{1}{3\bar{v}}\theta\left(x_{l}+\frac{3}{2}\gamma\bar{v}t,h^{2}t\right)h^{2}+O\left(h^{3}\right),
\end{aligned}
\label{eq:weakly-nonlin-limit}
\end{equation}
where the solutions of $\xi\left(z,\tau\right)$ and $\zeta\left(z,\tau\right)$
are given by the modulation equations \eqref{eq:asymp_O2} and the
solutions of $\rho\left(z,\tau\right)$ and $\theta\left(z,\tau\right)$
are given by the modulation equations \eqref{eq:asymp-coupled}. We
have the following theorem to ensure the convergence of the discrete
solution of the small-scale equation \eqref{eq:fluc-eqn} to the period-three
oscillations.
\begin{thm}
\label{thm:estim_peri3}Let $\rho\left(z,\tau\right)$ and $\theta\left(z,\tau\right)$
be smooth periodic solutions of \eqref{eq:asymp-coupled}, and $v_{l}\left(t\right),l=1,\cdots,L$
be the discrete solution of \eqref{eq:fluc-eqn}. Suppose that $v_{l}$
starts with the initial data consistent with $\rho_{0}\left(z\right)=\rho\left(z,0\right)$,
$\theta_{0}\left(z\right)=\theta\left(z,0\right)$ 
\begin{equation}
v_{l}\left(0\right)=\bar{v}+\cos\left(\frac{2\pi}{3}l\right)\rho_{0}\left(lh,0\right)h+\frac{1}{3\bar{v}}\theta_{0}\left(lh\right)h^{2},\label{eq:init_peri3}
\end{equation}
with $h=1/L$, and $\tilde{v}_{l}$ is defined by the approximation
\begin{equation}
\tilde{v}_{l}\left(t\right)=\bar{v}+\cos\left(\frac{2\pi}{3}l\right)\rho\left(x_{l}+\frac{3}{2}\gamma\bar{v}t,h^{2}t\right)h+\frac{1}{3\bar{v}}\theta\left(x_{l}+\frac{3}{2}\gamma\bar{v}t,h^{2}t\right)h^{2}.\label{eq:soln_peri3}
\end{equation}
Then, we have the $L^{2}$-estimate for the error between $v_{l}$
and $\tilde{v}_{l}$ for all $t\in\left[0,T\right]$
\begin{equation}
\left[\frac{1}{L}\sum_{l=1}^{L}\left|v_{l}\left(t\right)-\tilde{v}_{l}\left(t\right)\right|^{2}\right]^{\frac{1}{2}}\leq C_{T}\frac{1}{L^{2}}.\label{eq:esti_peri3}
\end{equation}
\end{thm}

\begin{proof}
The asymptotic expansion in \eqref{eq:eqn-peri3} implies that the
approximate solution \eqref{eq:soln_peri3} satisfies
\begin{equation}
\frac{d\tilde{v}_{l}}{dt}=-\frac{\gamma}{h}\tilde{v}_{l+1}\left(\tilde{v}_{l+2}-\tilde{v}_{l-1}\right)+T_{l},\label{eq:eqn_apprx}
\end{equation}
where $T_{l}=O\left(h^{3}\right)$ is the higher-order residual. On
the other hand, $v_{l}=\bar{v}+v_{l}^{\prime}$ satisfies \eqref{eq:fluc-eqn}
and we can rewrite its right hand side as the flux term and a reaction
term as in \eqref{eq:eqn_peri2}
\begin{equation}
\frac{dv_{l}}{dt}=-\frac{\gamma}{h}v_{l+1}\left(v_{l+2}-v_{l-1}\right)=-\frac{\gamma}{h}\left(G_{l+\frac{1}{2}}-G_{l-\frac{1}{2}}\right)-\frac{\gamma}{2h}\left(v_{l+1}-v_{l}\right)\left(v_{l+2}-v_{l-1}\right),\label{eq:eqn_flux}
\end{equation}
where we define the multivariable flux function $G\left(v_{1},v_{2},v_{3}\right)=\frac{1}{2}\left(v_{1}v_{2}+v_{1}v_{3}+v_{2}v_{3}\right)$
with $G_{l+\frac{1}{2}}\coloneqq G\left(v_{l},v_{l+1},v_{l+2}\right)$.
Denote the error as $e_{l}=\tilde{v}_{l}-v_{l}$ and we suppose \emph{a
priori} that 
\begin{equation}
\max_{l}\left|e_{l}\right|\leq h,\label{eq:apriori}
\end{equation}
for $0\leq t\leq t_{1}$ with a small enough $t_{1}$.

Since $\tilde{v}_{l}$ and $v_{l}$ satisfy \eqref{eq:eqn_apprx}
and \eqref{eq:eqn_flux} respectively, we have that the error $e_{l}$
satisfies the following equation
\begin{equation}
\begin{aligned}\frac{de_{l}}{dt}= & -\gamma\frac{G\left(\tilde{\mathbf{v}}_{l}\right)-G\left(\mathbf{v}_{l}\right)}{h}-\gamma\frac{G\left(\tilde{\mathbf{v}}_{l-1}\right)-G\left(\mathbf{v}_{l-1}\right)}{h}\\
 & -\frac{\gamma}{2}\left[\frac{\left(\tilde{v}_{l+1}-\tilde{v}_{l}\right)\left(\tilde{v}_{l+2}-\tilde{v}_{l-1}\right)}{h}-\frac{\left(v_{l+1}-v_{l}\right)\left(v_{l+2}-v_{l-1}\right)}{h}\right]+T_{l}.
\end{aligned}
\label{eq:err_estim}
\end{equation}
where we define $\mathbf{v}_{l}\coloneqq\left(v_{l},v_{l+1},v_{l+2}\right)$.
In the first line of the above equation, we have
\[
\left|\frac{G\left(\tilde{\mathbf{v}}_{l}\right)-G\left(\mathbf{v}_{l}\right)}{h}\right|=\left|\nabla_{\mathbf{v}}G\left(\mathbf{\tilde{v}}_{l}\right)\cdot\left(\tilde{\mathbf{v}}_{l}-\mathbf{v}_{l}\right)\right|\leq C\left(\left|e_{l}\right|+\left|e_{l+1}\right|+\left|e_{l+2}\right|\right).
\]
For the second row of \eqref{eq:err_estim}, we have the estimate
\begin{align*}
 & \frac{\left(\tilde{v}_{l+1}-\tilde{v}_{l}\right)\left(\tilde{v}_{l+2}-\tilde{v}_{l-1}\right)}{h}-\frac{\left(v_{l+1}-v_{l}\right)\left(v_{l+2}-v_{l-1}\right)}{h}\\
\leq & \left(\tilde{e}_{l+1}-\tilde{e}_{l}\right)\frac{\tilde{v}_{l+2}-\tilde{v}_{l-1}}{h}+\frac{\tilde{v}_{l+1}-\tilde{v}_{l}}{h}\left(e_{l+2}-e_{l-1}\right)-\frac{e_{l+1}-e_{l}}{h}\left(e_{l+2}-e_{l-1}\right)\\
\leq & C\left(\left|e_{l-1}\right|+\left|e_{l}\right|+\left|e_{l+1}\right|+\left|e_{l+2}\right|\right).
\end{align*}
Above, we have $\left|\tilde{v}_{l+i}-\tilde{v}_{l-j}\right|\leq Ch$
from the solution \eqref{eq:soln_peri3}, and $\left|e_{l}\right|\leq h$
from the supposition \eqref{eq:apriori}. Therefore, we have the estimate
for the error in \eqref{eq:err_estim} as
\[
\frac{de_{l}}{dt}\leq C_{1}\left(\left|e_{l-1}\right|+\left|e_{l}\right|+\left|e_{l+1}\right|+\left|e_{l+2}\right|\right)+C_{2}h^{3}.
\]

By multiplying $e_{l}$ on both sides of the above equation and taking
summation about $l$, we obtain
\begin{equation}
\begin{aligned}\frac{1}{2}\frac{d}{dt}\sum_{l}e_{l}^{2}\leq & C_{1}\sum_{l}\left|e_{l}\right|\left(\left|e_{l-1}\right|+\left|e_{l}\right|+\left|e_{l+1}\right|+\left|e_{l+2}\right|\right)+C_{2}\left|e_{l}\right|h^{3}\\
\leq & C^{\prime}\sum_{l}e_{l}^{2}+h^{5}.
\end{aligned}
\label{eq:err_energy}
\end{equation}
Using Gronwall's inequality, we have first for $t\leq t_{1}$, there
is 
\[
\sum_{l}e_{l}^{2}\leq Ch^{5}\;\Rightarrow\;\max_{l}\left|e_{l}\right|\leq Ch^{\frac{5}{2}}.
\]
Therefore, the a\emph{ priori} assumption \eqref{eq:apriori} is justified
first for $t\leq t_{1}$ then can always be continuously extended
to a larger time interval. This leads to the error estimate for the
entire time interval $t\in\left[0,T\right]$ such that
\[
h\sum_{l}e_{l}^{2}\leq Ch^{4}=\frac{C}{L^{4}}.
\]
This gives the total error in \eqref{eq:esti_peri3}.
\end{proof}
\begin{rem*}
The result in Theorem~\ref{thm:estim_peri3} can be also applied to
the single layer L96 model \eqref{eq:l96_num} describing the transition
between the period-three oscillations with the non-oscillatory region.
\end{rem*}

\subsection{Numerical verification using the two-layer L96 model\protect\label{subsec:Numerical-verification}}

In the numerical illustrations of the two-layer L96 solutions of \eqref{eq:l96-multi},
we consider two initial values: i) zero large-scale state $u_{1}\left(0\right)=u_{2}\left(0\right)=0$
and continuous small-scale profile $v_{0}=0.3\mathrm{sech}^{2}\left(\frac{x}{2}\right)$;
and ii) jump discontinuity at the two large-scale states $u_{1}\left(0\right)=1$,
$u_{2}\left(0\right)=-1$ and zero small scale $v_{0}\equiv0$. The
same as the one-layer model case, we use the discretization points
$L=256$ and the 4th-order Runge-Kutta method is used for the time
integration. First, the time evolution of the large-scale states are
shown in Figure \ref{fig:Time-evolution-2layer}. The large scale
and small scale mean interacts nonlinearly in the starting time. It
is observed in both test cases, the energy exchanges between the large
and small scale states and finally mostly transfers to the small-scale
state $v$ when strong chaotic behaviors are developed.

\begin{figure}
\subfloat{\includegraphics[scale=0.28]{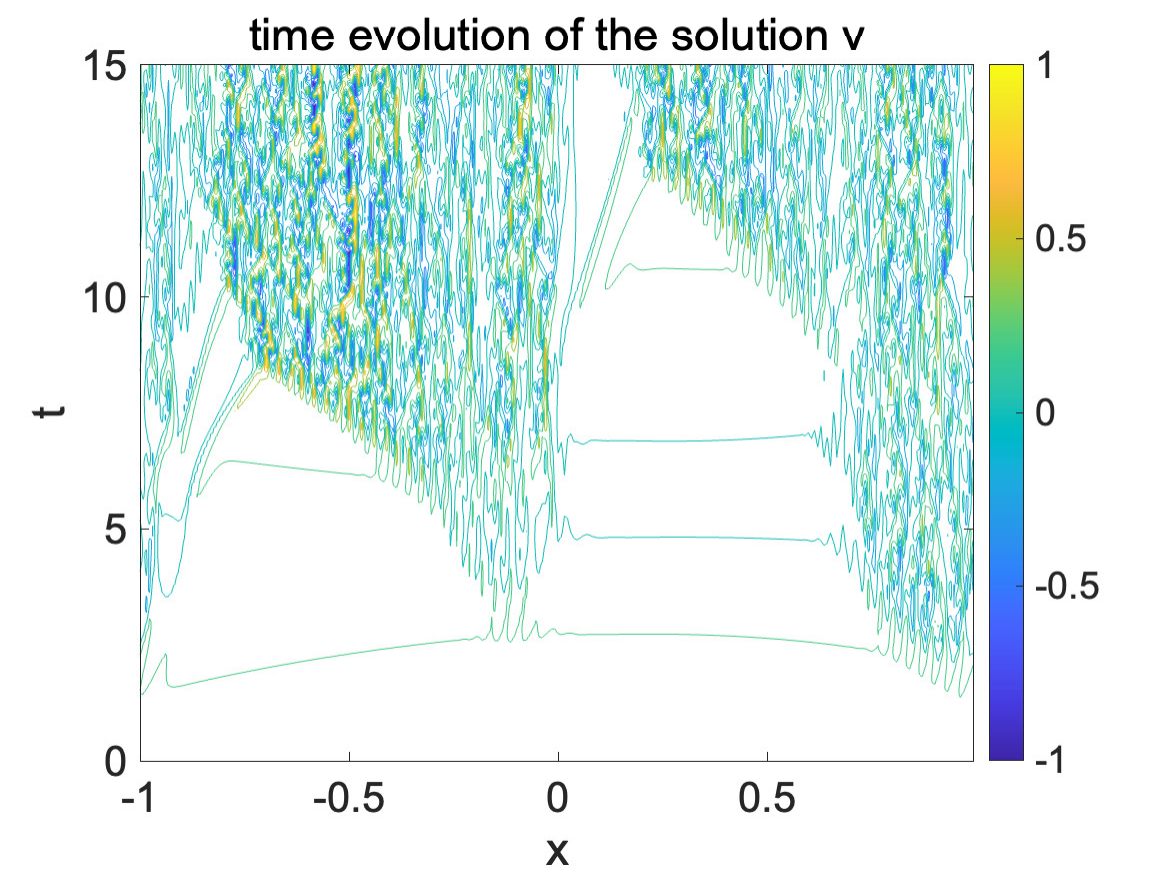}\includegraphics[scale=0.28]{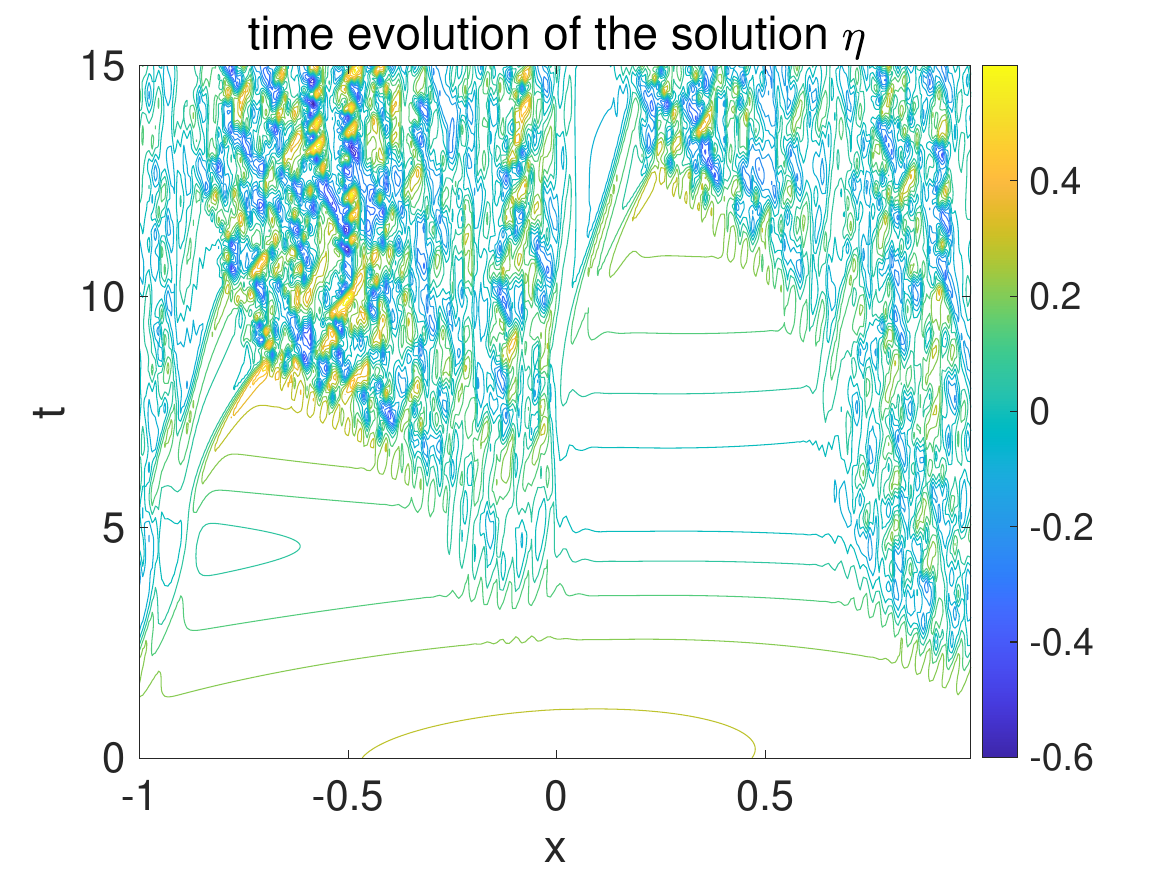}\includegraphics[scale=0.28]{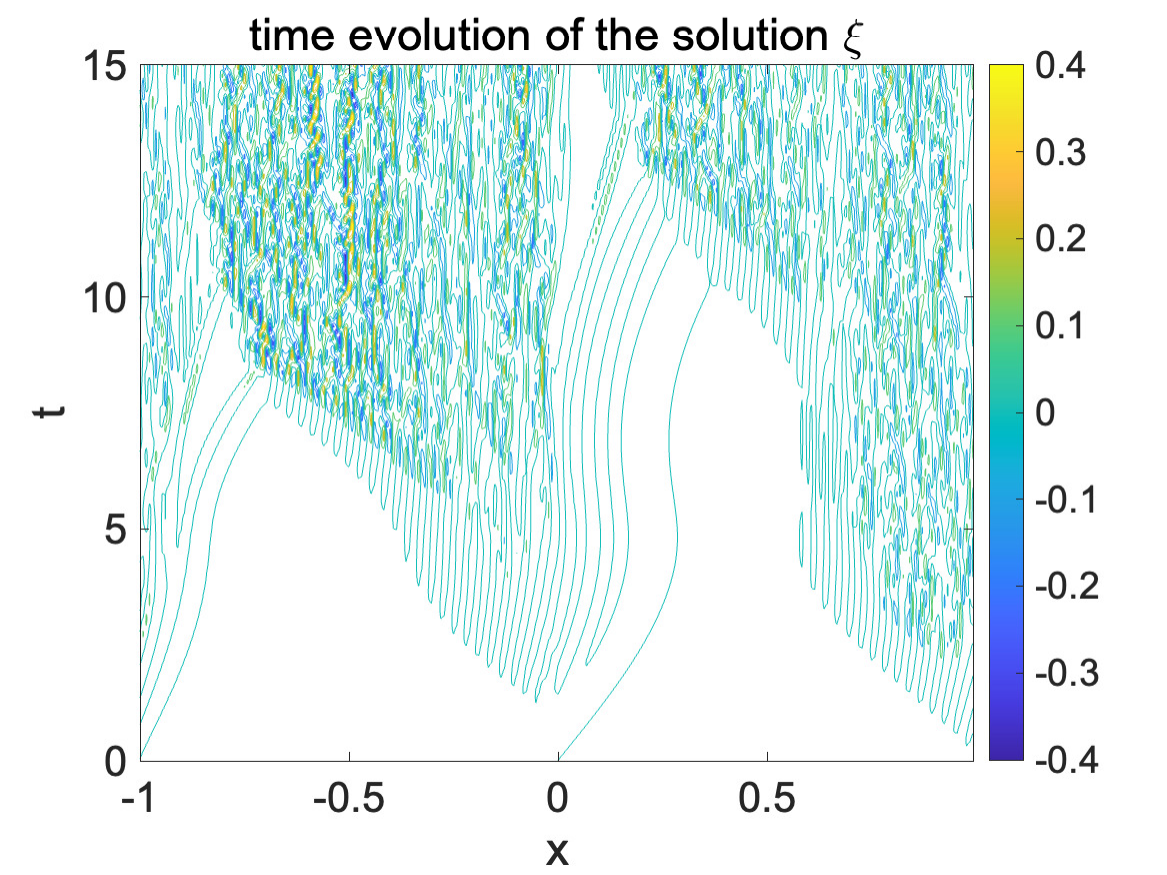}}

\subfloat{\includegraphics[scale=0.37]{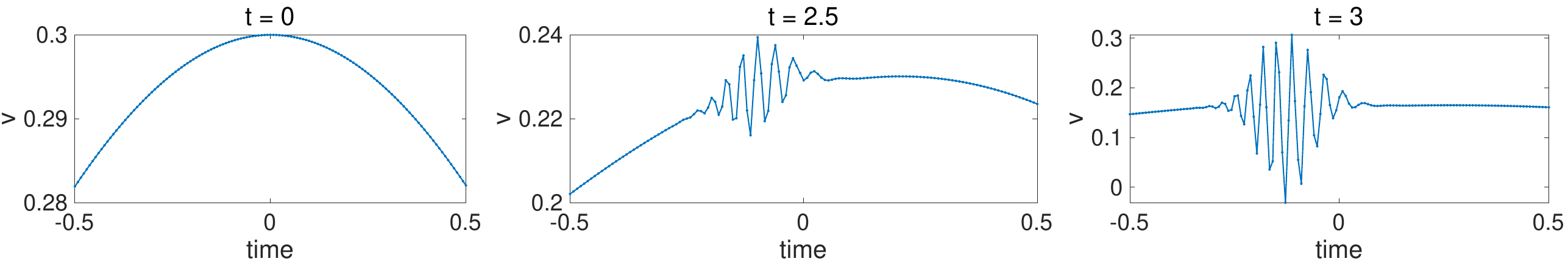}}

\caption{Development of oscillatory solutions in the small scales $v$ of the
two-layer L96 model with initial value $u_{1}\left(0\right)=u_{2}\left(0\right)=0$
and $v_{0}=0.3\mathrm{sech}^{2}\left(\frac{x}{2}\right)$.\protect\label{fig:Development-of-oscillatory1}}

\end{figure}

\begin{figure}
\subfloat{\includegraphics[scale=0.28]{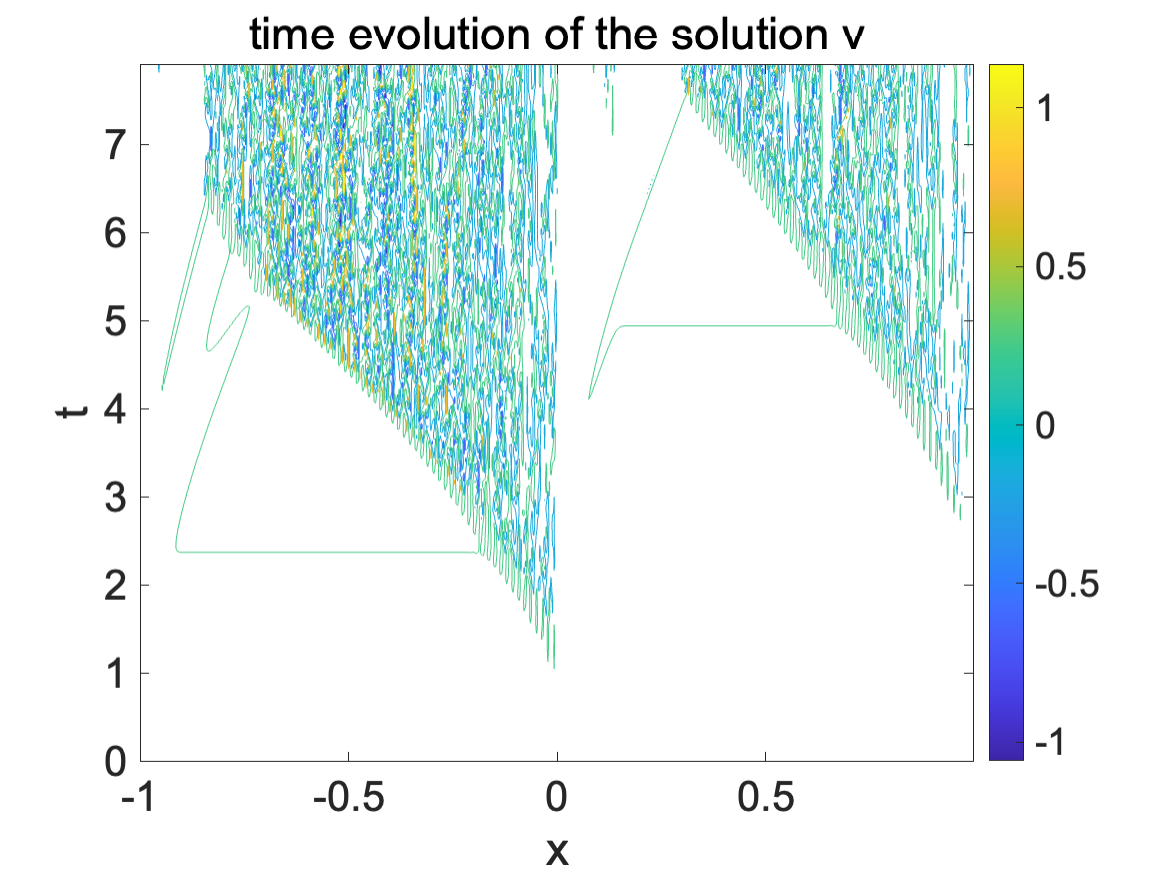}\includegraphics[scale=0.28]{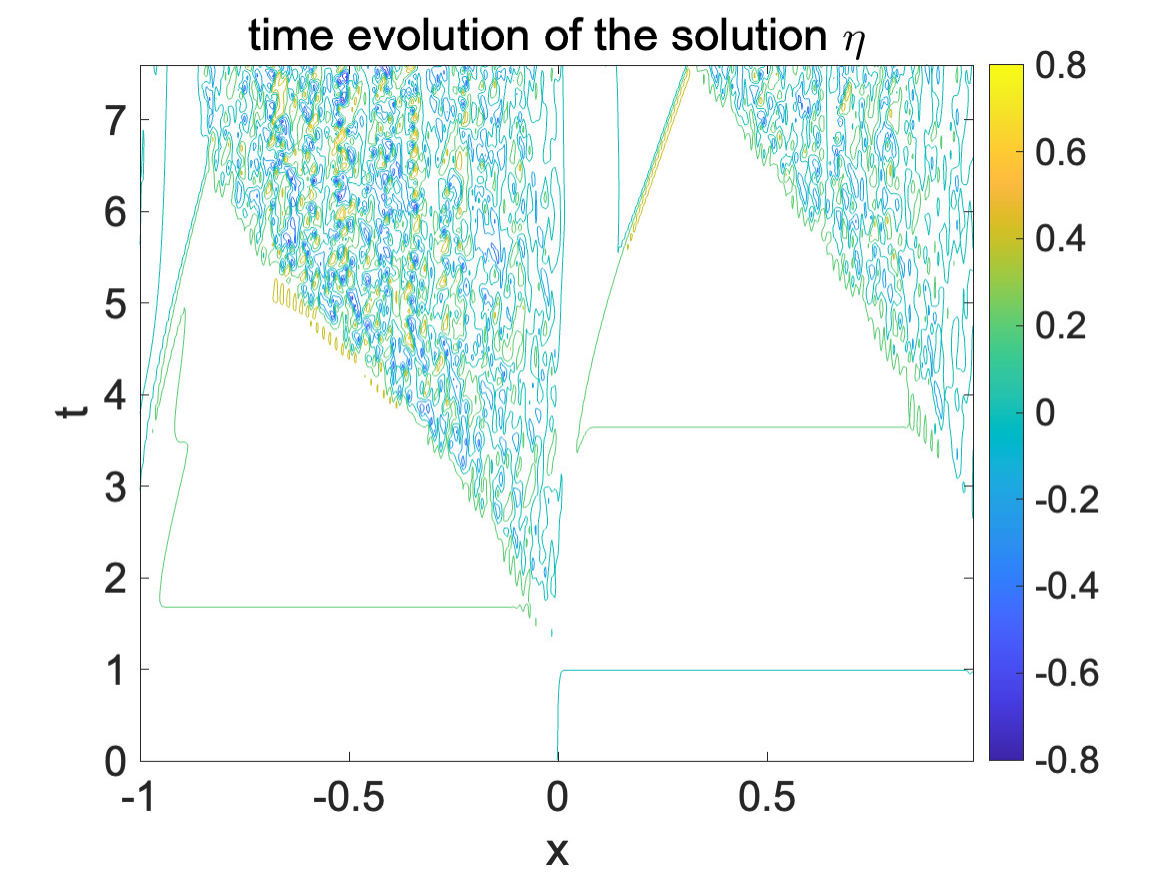}\includegraphics[scale=0.28]{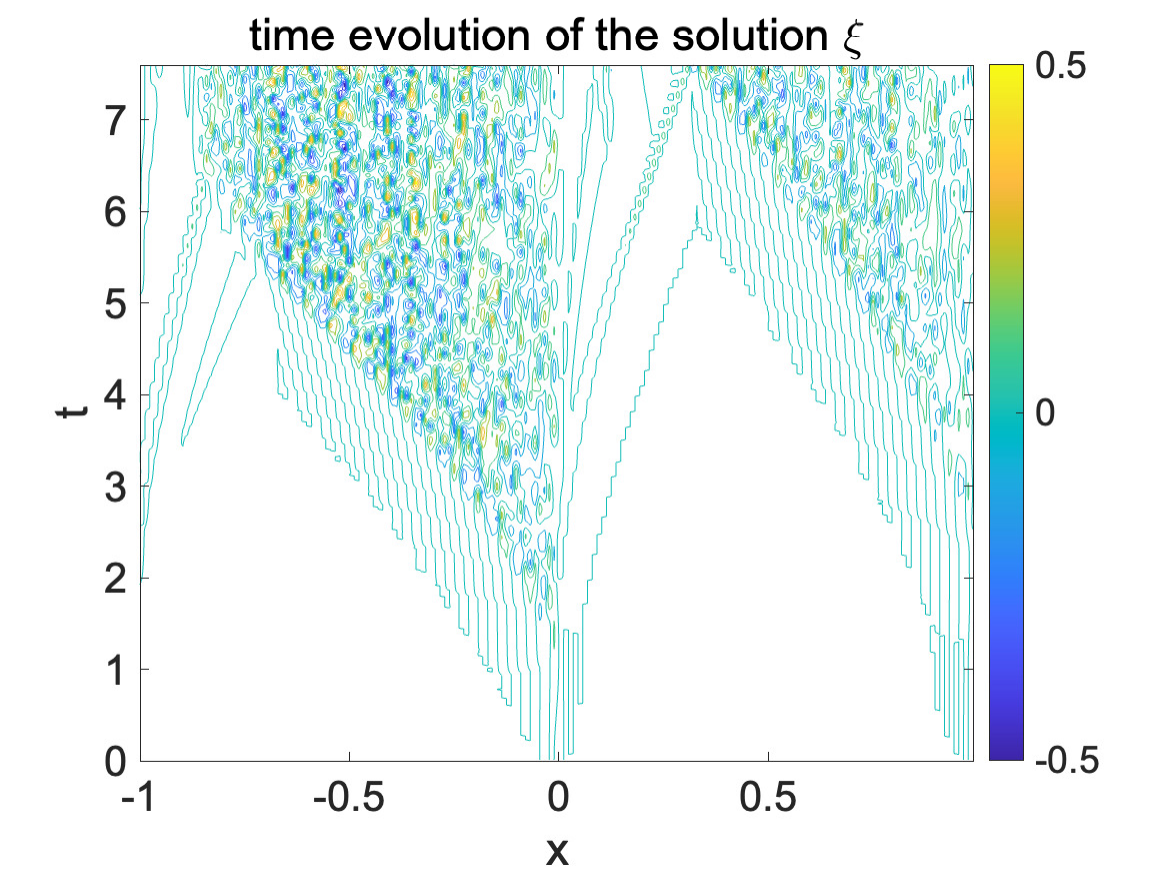}}

\subfloat{\includegraphics[scale=0.37]{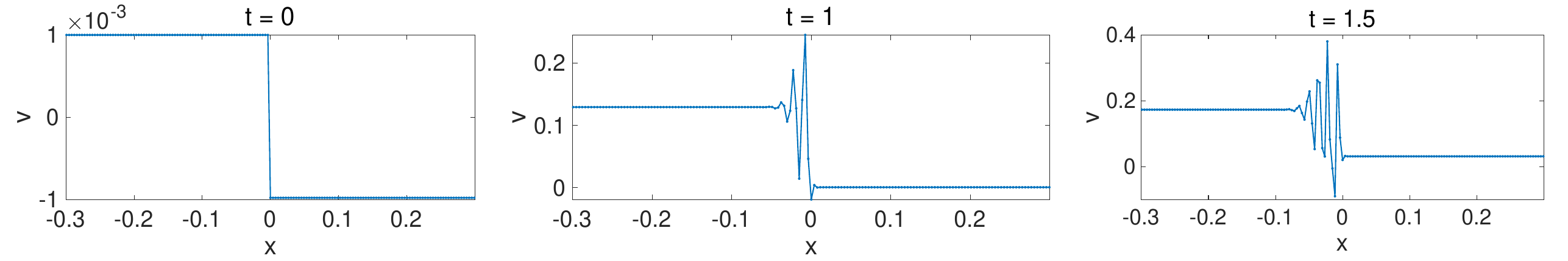}}

\caption{Development of oscillatory solutions in the small scales $v$ of the
two-layer L96 model with initial value $u_{1}\left(0\right)=1,u_{2}\left(0\right)=-1$
and $v_{0}=0$.\protect\label{fig:Development-of-oscillatory2}}
\end{figure}

The small-scale solutions in the two test cases are compared in Figure
\ref{fig:Development-of-oscillatory1} and Figure \ref{fig:Development-of-oscillatory2}.
From both initial values, oscillatory solutions in $v$ will be developed
near the point $x=0$ due to the different forcing $u_{1}$ and $u_{2}$
exerted on $v$ from the left and right sides of the domain according
to \eqref{eq:l96-multi}. It is observed the oscillations are moving
to the left side of the discontinuity due to the wave group velocity
$-c=-\frac{3}{2}\gamma\bar{v}$ consistent with the analysis in \eqref{eq:asymp_O0}.
We also approximate the period two solutions by $\eta=\frac{1}{3}\left(v_{l-1}+v_{l}+v_{l+1}\right)$
and $\xi=\frac{1}{3}\left(-\frac{1}{2}v_{l-1}+v_{l}-\frac{1}{2}v_{l+1}\right)$.
Notice that \eqref{eq:fluc-eqn} gives the leading-order small-scale
approximation since higher-order feedbacks from the large-scale states
also have contributions to the two-layer model \eqref{eq:l96-multi}
besides the leading-order terms in the large-scale equation \eqref{eq:large_contium}.
The combination of multiple approximation and numerical effects makes
it tricky to observe a clean period-three solution in the full two-layer
model simulations. Still, in the region where oscillatory solutions
start to develop, the three point averages $\eta,\xi$ demonstrate
smoother solution except some small fluctuations due to the higher-order
perturbations. This indicates the period-three behavior of the oscillatory
solution along its way to fully chaotic dynamics.

\section{Summary\protect\label{sec:Summary}}

We studied the development of chaotic behaviors commonly observed
in the simulations of the Lorenz 96 systems through the approach of
analyzing the convergence of discrete dispersive schemes. The final
chaotic solution can be explained as a competition between the stable
 solution and unstable oscillatory solutions developed during
the time evolution of the leading-order equations. Oscillations are
shown to arise in the discrete L96 models when discontinuity is developed
from the classical solution. We show that period-two oscillatory solutions
exist in the modulation equation, while period-three oscillations
may also occur in a weakly nonlinear analysis. Applying Strang's convergence
theorem to the regions of oscillatory solutions, we show that the
particular modulation equations and asymptotic weakly nonlinear approximations
characterize the period-two and period-three oscillations correspondingly.
The analytical results are supported by direct numerical simulations
in various parameter regimes of the one-layer and two-layer L96 models.
The ideas can be also generalized to study the more complicated interactions
of period-two and three oscillations that finally lead to the fully
turbulent dynamics.

\section*{Acknowledgements}

The research of D.Q. is partially supported by ONR grant N00014-24-1-2192
and NSF grant DMS-2407361. The research of J.-G. L. is partially supported
under the NSF grant DMS-2106988.

\bibliographystyle{plain}
\bibliography{refs_l96}

\begin{thebibliography}{10}

\bibitem{abramov2004quantifying}
Rafail~V Abramov and Andrew~J Majda.
\newblock Quantifying uncertainty for non-{G}aussian ensembles in complex
  systems.
\newblock {\em SIAM Journal on Scientific Computing}, 26(2):411--447, 2004.

\bibitem{arnold2013stochastic}
HM~Arnold, IM~Moroz, and TN~Palmer.
\newblock Stochastic parametrizations and model uncertainty in the {L}orenz '96
  system.
\newblock {\em Philosophical Transactions of the Royal Society A: Mathematical,
  Physical and Engineering Sciences}, 371(1991):20110479, 2013.

\bibitem{bedrossian2022regularity}
Jacob Bedrossian, Alex Blumenthal, and Sam Punshon-Smith.
\newblock A regularity method for lower bounds on the {L}yapunov exponent for
  stochastic differential equations.
\newblock {\em Inventiones mathematicae}, 227(2):429--516, 2022.

\bibitem{fatkullin2004computational}
Ibrahim Fatkullin and Eric Vanden-Eijnden.
\newblock A computational strategy for multiscale systems with applications to
  {L}orenz 96 model.
\newblock {\em Journal of Computational Physics}, 200(2):605--638, 2004.

\bibitem{goodman1988dispersive}
Jonathan Goodman and Peter~D Lax.
\newblock On dispersive difference schemes. {I}.
\newblock In {\em Selected Papers Volume I}, pages 545--567. Springer, 1988.

\bibitem{hou1991dispersive}
Thomas~Y Hou and Peter~D Lax.
\newblock Dispersive approximations in fluid dynamics.
\newblock In {\em Selected Papers Volume I}, pages 568--607. Springer, 1991.

\bibitem{karimi2010extensive}
Alireza Karimi and Mark~R Paul.
\newblock Extensive chaos in the {L}orenz-96 model.
\newblock {\em Chaos: An interdisciplinary journal of nonlinear science},
  20(4), 2010.

\bibitem{lax1986dispersive}
Peter~D Lax.
\newblock On dispersive difference schemes.
\newblock {\em Physica D: Nonlinear Phenomena}, 18(1-3):250--254, 1986.

\bibitem{lax1983small}
Peter~D Lax and C~David Levermore.
\newblock The small dispersion limit of the {K}orteweg-de {V}ries equation. i.
\newblock In {\em Selected Papers Volume I}, pages 463--500. Springer, 1983.

\bibitem{levermore1996large}
C~David Levermore and Jian-Guo Liu.
\newblock Large oscillations arising in a dispersive numerical scheme.
\newblock {\em Physica D: Nonlinear Phenomena}, 99(2-3):191--216, 1996.

\bibitem{liu1993stability}
Jian-Guo Liu and Zhou~Ping Xin.
\newblock L$^1$-stability of stationary discrete shocks.
\newblock {\em Mathematics of computation}, 60(201):233--244, 1993.

\bibitem{liu1993nonlinear}
Jian~Guo Liu and Zhouping Xin.
\newblock Nonlinear stability of discrete shocks for systems of conservation
  laws.
\newblock {\em Archive for rational mechanics and analysis}, 125:217--256,
  1993.

\bibitem{lorenz1998optimal}
Edward~N Lorenz and Kerry~A Emanuel.
\newblock Optimal sites for supplementary weather observations: Simulation with
  a small model.
\newblock {\em Journal of the Atmospheric Sciences}, 55(3):399--414, 1998.

\bibitem{lorenz1996proceedings}
EN~Lorenz.
\newblock Predictability: A problem partly solved.
\newblock {\em Proceedings of the Seminar on Predictability}, 1996.

\bibitem{majda2016introduction}
Andrew~J Majda.
\newblock {\em Introduction to turbulent dynamical systems in complex systems}.
\newblock Springer, 2016.

\bibitem{majda2018strategies}
Andrew~J Majda and Di~Qi.
\newblock Strategies for reduced-order models for predicting the statistical
  responses and uncertainty quantification in complex turbulent dynamical
  systems.
\newblock {\em SIAM Review}, 60(3):491--549, 2018.

\bibitem{majda2019linear}
Andrew~J Majda and Di~Qi.
\newblock Linear and nonlinear statistical response theories with prototype
  applications to sensitivity analysis and statistical control of complex
  turbulent dynamical systems.
\newblock {\em Chaos: An Interdisciplinary Journal of Nonlinear Science},
  29(10), 2019.

\bibitem{majda2000remarkable}
Andrew~J Majda and Ilya Timofeyev.
\newblock Remarkable statistical behavior for truncated {B}urgers--{H}opf
  dynamics.
\newblock {\em Proceedings of the National Academy of Sciences},
  97(23):12413--12417, 2000.

\bibitem{olbers2001gallery}
Dirk Olbers.
\newblock A gallery of simple models from climate physics.
\newblock In {\em Stochastic Climate Models}, pages 3--63. Springer, 2001.

\bibitem{orrell2003model}
David Orrell.
\newblock Model error and predictability over different timescales in the
  {L}orenz '96 systems.
\newblock {\em Journal of the atmospheric sciences}, 60(17):2219--2228, 2003.

\bibitem{qi2023high}
Di~Qi and Jian-Guo Liu.
\newblock High-order moment closure models with random batch method for
  efficient computation of multiscale turbulent systems.
\newblock {\em Chaos: An Interdisciplinary Journal of Nonlinear Science},
  33(10), 2023.

\bibitem{qi2023random}
Di~Qi and Jian-Guo Liu.
\newblock A random batch method for efficient ensemble forecasts of multiscale
  turbulent systems.
\newblock {\em Chaos: An Interdisciplinary Journal of Nonlinear Science},
  33(2), 2023.

\bibitem{strang1964accurate}
Gilbert Strang.
\newblock Accurate partial difference methods: {II}. non-linear problems.
\newblock {\em Numerische Mathematik}, 6(1):37--46, 1964.

\bibitem{stuart1998dynamical}
Andrew Stuart and Anthony~R Humphries.
\newblock {\em Dynamical systems and numerical analysis}, volume~2.
\newblock Cambridge University Press, 1998.

\bibitem{wilks2005effects}
Daniel~S Wilks.
\newblock Effects of stochastic parametrizations in the {L}orenz '96 system.
\newblock {\em Quarterly Journal of the Royal Meteorological Society: A journal
  of the atmospheric sciences, applied meteorology and physical oceanography},
  131(606):389--407, 2005.

\bibitem{xin1992linearized}
Zhouping Xin.
\newblock On the linearized stability of viscous shock profiles for systems of
  conservation laws.
\newblock {\em Journal of differential equations}, 100(1):119--136, 1992.

\end{thebibliography}

\end{document}